\documentclass[12pt]{article}
\usepackage{amsmath}
\usepackage{mathtools} 
\usepackage[T1]{fontenc}
\usepackage{amssymb}
\usepackage{amsfonts}
\usepackage{amsthm}
\usepackage{mathrsfs}
\usepackage[all]{xy}
\usepackage{tensor}
\usepackage{comment}
\usepackage{bm}
\usepackage{bbm}
\usepackage[normalem]{ulem}
\usepackage[usenames,dvipsnames]{xcolor}
\usepackage[colorlinks=true, citecolor=Blue, linkcolor=Blue]{hyperref}
\usepackage{ragged2e}
\usepackage{tikz}
    \usetikzlibrary{decorations.markings}
    \usetikzlibrary{cd} 

\usepackage{setspace}
\usepackage[labelfont={small, bf, sf}, textfont={small,sf}]{caption}
\numberwithin{equation}{section} 
\usepackage{authblk}
\usepackage[in]{fullpage}
\usepackage{cite}

\newcommand{\op}{\mathcal{O}}
\newcommand{\vep}{\varepsilon} 
\newcommand{\lie}{\pounds}

\newcommand{\tone}{\text{I}}
\newcommand{\ttwo}{\text{II}}
\newcommand{\tthr}{\text{III}}

\newcommand{\alg}{\mathcal{A}}

\newcommand{\defeq}{\coloneqq}
\DeclareMathOperator{\aut}{Aut}
\DeclareMathOperator{\cnt}{Cnt}
\DeclareMathOperator{\Int}{Int}
\newcommand{\agrav}{\widetilde{\alg}}
\newcommand{\rinj}{\mathcal{R}}
\newcommand{\algm}{\mathcal{M}}
\DeclareMathOperator{\Ad}{Ad}
\DeclareMathOperator{\out}{Out}
\DeclareMathOperator{\uni}{API}
\DeclareMathOperator{\Mod}{mod}
\newcommand{\hobs}{\hat{\vep}}
\newcommand{\agravss}{\widetilde{\alg}_0}

\newcommand{\Tovw}{\mc{T}}
\newcommand{\Sovw}{\mc{S}}
\newcommand{\tp}{\hat{t}}
\newcommand{\aqft}{\alg_\text{QFT}}
\newcommand{\aext}{\alg_\text{ext}}

\newcommand{\mc}{\mathcal}

\newcommand{\msf}{\mathsf}
\newcommand{\wh}{\widehat}
\newcommand{\tr}{\operatorname{Tr}}

\newcommand{\hs}{\mathcal{H}}

\newcommand{\beq}{\begin{equation}}
\newcommand{\eeq}{\end{equation}}
\newcommand{\bes}{\begin{subequations}}
\newcommand{\ees}{\end{subequations}}
\newcommand{\bea}{\begin{eqnarray}}
\newcommand{\eea}{\end{eqnarray}}

\title{An intrinsic cosmological observer}

\author[1,2]{Antony J. Speranza\thanks{asperanz@gmail.com}}

\affil[1]{\small \it Institute for Theoretical Physics, University of Amsterdam, Science Park 904, 1098 XH, Amsterdam, The Netherlands}
\affil[2]{\small \it Department of Physics, University of Illinois, Urbana-Champaign, Urbana IL 61801, USA}

\date{April 10, 2025}

\begin{document}

\maketitle

\begin{abstract}

There has been much recent interest in the necessity of an observer degree of freedom in the description of local algebras in semiclassical gravity.  In this work, we describe an example where the observer can be constructed intrinsically from the quantum fields.  This construction involves the slow-roll inflation example recently analyzed by Chen and Penington, in which the gauge-invariant gravitational algebra arises from marginalizing over modular flow in a de Sitter static patch.  We relate this procedure to the Connes-Takesaki theory of the flow of weights for type III von Neumann algebras, and further show that the resulting gravitational algebra can naturally be presented as a crossed product.  This leads to a decomposition of the gravitational algebra into quantum field and observer degrees of freedom, with different choices of observer being related to changes in 
a quantum reference frame for the algebra.  We also connect this example to other constructions of type II algebras in semiclassical gravity, and argue they all share the feature of being the result of gauging modular flow.  The arguments in this 
work involve various properties of automorphism
groups of hyperfinite factors, and so in an appendix we 
review the structure of these  groups, which
may be of independent interest for further investigations
into von Neumann algebras in quantum gravity.

\end{abstract}

\flushbottom

\newpage

\tableofcontents

\section{Introduction}

Entanglement entropies for subregions in quantum field 
theory suffer from well-known UV divergences 
coming from the highly entangled structure 
of the vacuum at short distances.  
This leads to various challenges when working 
with QFT entanglement entropies, since they are quite sensitive
to the choice of regulator, and extracting regulator-independent
quantities can be a delicate process.  It also
prevents one from using rigorous algebraic techniques,
such as the Haag-Kastler approach involving von Neumann
algebras for local subregions \cite{Haag1964, Haag1992}, 
to define entanglement entropies,
thereby
also precluding  the 
powerful information-theoretic tools that come with 
such an algebraic formulation 
\cite{Petz1993}.
The algebraic approach instead forces one to 
focus on UV-finite quantities such as relative entropies
or vacuum-subtracted entropies computed with 
specific regularization schemes 
\cite{Casini:2006ws, Casini:2008wt, Casini:2015woa,
Dutta:2019gen, 
Kudler-Flam:2023hkl}.

The situation in quantum gravity is expected to be 
much better, where the entropy associated with the 
exterior of a black hole, for example, is given by the generalized 
entropy $S_\text{gen} = \frac{A}{4G} + S_\text{out}$.  
This is a finite quantity whose leading contribution
is given by the Bekenstein-Hawking entropy $\frac{A}{4G}$ \cite{Bekenstein1972, Bekenstein1973a, Hawking:1975vcx},
and $S_\text{out}$ is the entropy of degrees of freedom
in the region outside the black hole horizon.  By interpreting
$S_\text{out}$ as the entanglement entropy of quantum
fields restricted to the black hole exterior, one can 
argue that the generalized entropy is UV-finite,
with the UV-divergence in the entanglement entropy canceling
against loop effects that renormalize the gravitational
coupling constant $G$
\cite{Susskind:1994sm, Jacobson:1994iw, Larsen:1995ax,
Cooperman:2013iqr, Solodukhin:2011gn,Bousso:2015mna}. 

Recently, there has been much interest in the definition of 
entropy in semiclassical gravity, in which one takes the limit of 
small gravitational coupling, $G\rightarrow 0$.  In this 
limit the generalized entropy diverges since $G$ appears in the 
denominator of the Bekenstein-Hawking term, and this is the expected
behavior when gravity decouples, since this limit admits a description
in terms of quantum field theory in curved spacetime.  However,
a number of recent papers have argued that this description
is modified when properly accounting for diffeomorphism
constraints coming from the interacting theory
\cite{Witten2021, Chandrasekaran2022a, Chandrasekaran2022b,
Penington:2023dql, Kolchmeyer:2023gwa,
Jensen2023, Kudler-Flam:2023qfl, Ali:2024jkx, Faulkner:2024gst, 
Kudler-Flam:2024psh, Chen2024, Penington:2024sum}.  This modification
requires that one explicitly include certain global gravitational
degrees of freedom, such as the ADM mass in a black hole background, 
and that operators localized to a region of spacetime 
be appropriately dressed to these global degrees of freedom.  
This has a dramatic effect on the local algebras.  
In the language of von Neumann algebras,  the 
undressed operators form an 
algebra of type $\tthr_1$,
while the dressed algebras are type $\ttwo$.

Type $\ttwo$ algebras differ from their type $\tthr$ counterparts in
that every state can be described by a density matrix, and hence
be associated with a 
renormalized notion of entropy.  These properties follow from 
the existence of a {\it semifinite trace} defined on the algebra.  
Given a von Neumann algebra $\alg$,  i.e.\ a weakly closed 
algebra of bounded operators acting on a Hilbert space, 
a trace is a linear functional $\tau$  
on the algebra---or rather, on a subalgebra $\mathfrak{m}\subset \alg$ 
of  operators, known as the {\it definition ideal}, 
for which $\tau$ is finite---satisfying the cyclic property
\beq
\tau(\msf{ab}) = \tau(\msf{ba}), \qquad \msf{a},\msf{b}\in\mathfrak{m}.
\eeq
In the familiar case where $\alg$ is the  type $\tone_\infty$ 
algebra of all bounded linear operators acting on an infinite-dimensional
Hilbert space, the trace is given by the usual formula summing 
over the expectation values in an orthonormal basis.  In this 
case, $\mathfrak{m}$ consists of the trace class operators.  More generally, 
one says that a trace is semifinite if $\mathfrak{m}$ is weakly dense 
in $\alg$.  One of the defining properties of type $\tthr$ algebras
is that they do not possess a (normal) 
semifinite  trace, whereas type $\tone$ and 
 $\ttwo$ algebras do \cite[Section V.2]{TakesakiI}.  

The density matrix $\rho$ for a state $\omega$ is then defined as the positive 
operator affiliated with $\alg$ that reproduces expectation
values when inserted into the trace,
\beq
\omega(\msf{a}) = \tau(\rho\msf{a}).
\eeq
One then defines an entropy for the state in terms of $\rho$ by 
the formula
\beq \label{eqn:Srho}
S(\rho) = -\tau(\rho\log\rho).
\eeq
For type $\tone$ factors where the trace is normalized to  
$1$ on the minimal projections, this is the familiar von Neumann
entropy of the density matrix, which is positive and bounded above 
by the dimension of the Hilbert space on which $\alg$ acts irreducibly.  
For type $\ttwo$ factors, the entropy defined by (\ref{eqn:Srho})
is known as the {\it Segal entropy} \cite{Segal1960}\cite[Chapter 7]{Petz1993},
and has somewhat different properties.  It is no longer required 
to be positive, and ranges either from $-\infty$ to $\infty$ 
when $\alg$ is type $\ttwo_\infty$, or from $-\infty$ to $0$ 
when $\alg$ is type $\ttwo_1$, where the trace is normalized in the 
latter case by $\tau(\mathbbm{1}) = 1$.  Although a negative 
entropy  might sound strange, it simply follows from 
the appropriate interpretation of the Segal entropy as 
an entropy difference
from a reference state.  This interpretation is related to the fact 
that for type $\ttwo$ algebras, the trace $\tau$ is not the usual
Hilbert space trace, but rather something more like an infinitely 
rescaled version of it.  This renormalization of the trace results in
a shift in the entropy, and allows for negative values to be 
attained.  

Returning to the discussion of semiclassical quantum gravity,
we see the statement that the algebra is type $\ttwo$ translates to 
the statement that entropy differences are well defined in the $G\rightarrow 0$
limit of quantum gravity.  Furthermore, calculations of the perturbative 
backreaction of states of the quantum fields on the geometry have 
demonstrated that these entropy differences match onto differences in
the generalized entropy \cite{Chandrasekaran2022a,
Chandrasekaran2022b, Jensen2023}.  
Hence, even at infinitesimally weak gravitational
coupling where the full generalized entropy diverges, a well-defined notion
of vacuum-subtracted generalized entropy exists, and can be computed 
using algebraic techniques.  

The way in which the type $\ttwo$ structure was identified for the
gravitational algebras was through an algebraic construction
known as the modular crossed product.  This construction
enlarges a given algebra $\alg$ by an operator that generates
modular flow of a state on $\alg$, and for type $\tthr$ algebras 
it is known via the work of Takesaki to result in a type $\ttwo$ 
algebra \cite{Takesaki1973}
\cite[Chapter XII]{TakesakiII}.  While the mathematics
underlying in the construction is clear, 
there remain some questions regarding the intuitive explanation
behind the emergence of a renormalized trace and entropy.  
Two points in particular stand out.  The first is that although
it is straightforward to check that the crossed product algebra
possesses a trace via computations involving Tomita-Takesaki theory,
it is not immediately clear what the underlying mechanism is 
that leads to this trace, and in particular, whether it relies on the 
specific details in each construction, or if it is a more generic
feature.  The second question involves the interpretation of the additional
operator that is added to the algebra in the crossed product constructions.
This operator has different interpretations in different contexts:
for constructions involving black holes, it is interpreted as an asymptotic
charge such as the ADM Hamiltonian, while for the static patch of de Sitter,
the additional operator is an extra degree of freedom associated with 
an observer.  The de Sitter static patch example is particularly vexing, since
one might have expected that the quantum fields themselves provide the 
full collection physical observables, so it feels somewhat unnatural
to  include an additional degree of freedom by hand.  

In this work, we will address these two points.  
The first question relates to the underlying reason for the emergence
of a trace in the gravitational algebra constructions.  
Our proposal for how to understand this result is that the gravitational
algebras arise from gauging modular flow on a kinematical type $\tthr$
algebra $\alg$.  The justification for gauging modular flow is the 
so-called {\it geometric modular flow conjecture} proposed in 
\cite{Jensen2023}.  The intuition for this conjecture is that 
modular flow for the type $\tthr_1$ algebras describing operators localized
to a subregion in quantum field theory should look like a boost close
to the entangling surface.  This follows from the expectation that 
on the one hand, all states in quantum field theory should approach the vacuum at 
short distances, while on the other hand, all entangling surfaces look
locally like the bifurcation surface of Rindler space at short 
distances.  According to the Bisognano-Wichmann theorem and the related Unruh
effect \cite{Bisognano1975, Unruh:1976db}, 
the vacuum modular Hamiltonian of Rindler space is precisely
the boost generator in Minkowski space.  The precise statement 
of the geometric modular flow conjecture is then that given 
a flow that acts as a diffeomorphism on a Cauchy slice and approaches
a constant-surface-gravity boost at the entangling surface, there 
exists a normal, semifinite weight on the algebra $\alg$ whose
modular Hamiltonian generates the given flow. For some works
investigating aspects of this conjecture,
see \cite{Sorce:2024zme, Caminiti:2025hjq}.

Since diffeomorphisms are gauged in gravitational theories,
the geometric modular flow conjecture then suggests that 
the appropriate dressed algebra arises from gauging 
modular flow.  In practice, this means that starting from
a type $\tthr_1$ kinematical algebra of quantum fields 
in a causally complete subregion of a background spacetime,
the dressed algebra consists of operators that commute
with the modular flow of the weight whose modular
Hamiltonian generates the boost.  
The collection of operators that commute with modular
flow of a weight $\omega$ on an algebra is called 
the {\it centralizer} of the weight, denoted $\alg_\omega$,
and in many cases such centralizers possess a trace.
The reason for this is fairly intuitive: one can think of modular 
flow as being generated by the density matrix $\rho_\omega$ for the weight,
and so
the centralizer then consists of operators that commute with the 
density matrix.  For such operators, the original weight $\omega$
defines a trace.  We can verify this statement
explicitly in the type $\tone$ and $\ttwo$ case, where the 
density matrix is well-defined.  In those cases, we have
for $\msf{a},\msf{b}$ in the centralizer $\alg_\omega$,
\beq
\omega(\msf{ab}) = \tr\left(\rho_\omega \msf{ab}\right)
=\tr \left(\msf{a}\rho_\omega\msf{b}\right)
=\tr \left(\rho_\omega\msf{ba}\right)
= \omega(\msf{ba}).
\eeq

Crucially, the fact that $\omega$ defines a trace 
on its centralizer holds also in the type $\tthr$ 
case \cite[Theorem
VIII.2.6]{TakesakiII}, 
even though the above argument involving
density matrices is not applicable.
This statement is almost enough to conclude the 
centralizer possesses a trace; however, there is a 
subtlety when working with proper unbounded weights, since
they might not define finite expectation values on 
their centralizer; see the discussion in section \ref{sec:intweight}.
The cases of interest for recent works on gravitational
algebras, however, all are examples involving centralizers
with a well-defined trace.  In particular, both 
the crossed product construction and the observer
in the static patch of de Sitter are examples of 
algebras that arise as centralizers of weights,
as explained explicitly in section \ref{sec:vacds}.
This then sheds some light on why semiclassical gravitational
algebras possess renormalized traces and entropies: both
arise as a consequence of gauging modular flow.  

The second point we address in this work is the interpretation
of the observer operator appearing in certain constructions 
of gravitational algebras.  This operator 
features prominently in the Chandrasekaran-Longo-Penington-Witten
(CLPW) algebra for the static
patch of de Sitter \cite{Chandrasekaran2022a}, and was necessary for a seemingly
technical reason: the modular flow for the static
patch of de Sitter is ergodic, meaning the centralizer consists
only of multiples of the identity.  By enlarging the 
algebra for the static patch by an observer with nontrivial
energy, CLPW found a type $\ttwo_1$ algebra associated 
with the centralizer, and with it a nontrivial notion of 
entropy.  However, one might have expected that the observer should
be constructed intrinsically from the quantum fields.

A step in this direction was provided by Chen and Penington
\cite{Chen2024}, who considered a modification of the CLPW construction
involving a slowly rolling inflaton scalar field.  The potential
for the scalar was chosen so that the field
eternally decays to smaller values, and hence it can be used 
to define a clock with respect to which operators 
can be dressed.  They argued directly that the centralizer
for the natural weight associated with this potential
is nontrivial, and results in a type $\ttwo_\infty$
gravitational algebra.  This argument
involves an averaging procedure over
modular flow, and as explained in the present work
in section \ref{sec:intweight}, 
 such a procedure is directly related 
to the Connes-Takesaki theory of integrable weights
\cite{Connes1977}.\footnote{The connection
between inflationary and crossed-product algebras 
and the Connes-Takesaki theory has also been
emphasized in \cite{Gomez:2023tkr, Gomez:2023jbg,
Gomez:2024kuy}.  However, we differ in some of our
conclusions, in part because the above-mentioned
works contain some erroneous claims about the 
Connes-Takesaki classification theorem and properties
of centralizers.  } 
For Chen and Penington, the existence of 
an observer is implicitly associated with the clock
that the rolling scalar field provides, but they 
do not define an explicit operator that plays the 
role of the observer Hamiltonian.  

A central result of this paper is to show how such
an operator can be constructed, thereby providing
an intrinsic notion of observer 
constructed from the quantum fields.  We make this 
identification by showing that the inflationary 
algebra of Chen and Penington has a natural representation
as a crossed product algebra.  As explained
in section \ref{sec:crossprod}, the key feature 
in exhibiting this crossed product is the 
existence of a preferred family of automorphisms that rescale the 
trace defined on the inflationary algebra.  These automorphisms
arise from the shift symmetry of the scalar field.  The 
shift-symmetric operators form a type $\tthr_1$ subalgebra
$\agrav_0$,
and the full inflationary algebra has the structure 
of a modular crossed product of $\agrav_0$.  The observer
Hamiltonian is simply the additional operator that must 
be added to $\agrav_0$ to generate the full inflationary
algebra.  This choice of additional operator is not canonical,
and we argue that different choices are related to crossed
product descriptions by different weights.  This ambiguity
in the choice of observer has an interpretation as 
a choice of quantum reference frame for the description
of the algebra.  We give a comparison to other recent works
on quantum reference frames and gravitational algebras 
in section \ref{sec:qrf}

From this example, we conclude that there are two separate
effects occurring in the construction of gravitational
algebras.  The existence of a trace and renormalized entropies
is not directly related to the inclusion of an observer; rather,
it is the result of gauging modular flow.  The existence of an 
observer operator is additional structure associated with the 
algebra coming from representing it as a crossed product.  

In section \ref{sec:vacds}, we re-analyze the CLPW construction
of an observer in the dS static patch from the perspective 
of gauging modular flow.  We show that the algebra 
in this case is once again the centralizer 
of an integrable weight, and the Connes-Takesaki
classification gives an explanation for 
why a type $\ttwo_1$ algebra arises in this case.  
A possible puzzle arises in this 
context since the type $\ttwo_1$ centralizer 
does not admit a description as a modular crossed product,
nor does it have a type $\tthr_1$ subalgebra 
associated with the quantum field operators.  
We describe a possible resolution,
in that one should instead focus on the existence
of time operators in the kinematical algebra, and 
define the observer Hamiltonian as a canonical 
conjugate in the gravitational algebra to this operator.  
The connection between observers and time operators was 
also a crucial aspect of the construction of an 
observer in holography at large $N$ recently explored 
in \cite{Jensen:2024dnl}, and has also been
emphasized in cosmological setups in 
\cite{Gomez:2023wrq, Gomez:2023upk}.  
This picture points to a broader
interpretation for how to understand observers and time 
operators in gravitational algebras, 
and we discuss that the correct general description 
should be in terms of a subfactor inclusion $\agrav\subset
\alg$ of a semifinite gravitational algebra $\agrav$ inside
a larger type $\tthr_1$ algebra $\alg$ 
that includes  time operators.  

The paper is organized as follows.  In section \ref{sec:inflalg},
we describe the setup of the kinematical algebra for the 
slow-roll inflation gravitational algebra.  We work
with a simplified version of the model considered 
by Chen and Penington  by restricting to 
two-dimensional de Sitter space; 
this model contains the same essential feature as the more
realistic four dimensional model from \cite{Chen2024}, but 
has the advantage of being computationally simpler.  We derive
the Bunch-Davies weight for a scalar field with linear potential,
and argue in section \ref{sec:intweight} that it defines 
a dominant weight, in the terminology of the Connes-Takesaki
classification \cite{Connes1977}.  Then in section 
\ref{sec:crossprod}, we show that the resulting 
algebra has a canonical description as a crossed product
coming from the existence of a preferred trace-scaling automorphism.
Section \ref{sec:constr} gives a procedure for perturbatively
constructing operators in the centralizer of the Bunch-Davies
weight, as well as a construction of the intrinsic 
observer Hamiltonian present in the crossed-product description.
Section \ref{sec:time} discusses the relation 
between the crossed product description and an existence
of a time operator, and points out a small puzzle regarding  the 
appropriate identification of a time operator.  
Finally in section \ref{sec:vacds}, we analyze the 
CLPW construction involving an observer in the de Sitter 
static patch, and connect the resulting $\ttwo_1$ algebra 
to the Connes-Takesaki classification of integrable weights.
We conclude in section \ref{sec:discussion} with some 
open questions and ideas for future work.  In much 
of this work, we make use of certain properties of 
automorphism groups of hyperfinite factors.  Therefore
in appendix
\ref{sec:automorphism}, we review a number of facts about
the structure of these automorphism groups; this
summary 
may be useful for other investigations into von Neumann
algebras in semiclassical gravity and quantum field theory.

\paragraph{Notation.}
We will assume some familiarity with von Neumann algebras 
and properties of modular flows; we refer to 
\cite{Witten:2018zxz} for an accessible introduction.
We generally denote the modular automorphism
on an algebra by $\sigma_t$, but note that our convention
is $\sigma_t(\msf{a}) = e^{ith}\msf{a}e^{-ith}$,
where $\msf{a}$ is an element of a von Neumann algebra $\alg$,
and $h = -\log \Delta$ is the modular Hamiltonian of 
a weight.  This definition uses the opposite convention
for the sign of $t$ from most of the mathematics literature
(e.g.\ \cite{TakesakiII}), and this results in some 
different sign conventions in certain formulas.  We use angle
brackets to denote the von Neumann algebra generated
by the listed operators or algebras, so, for example
$\langle \mathcal{A}, \mathcal{B}\rangle = \mathcal{A}\vee
\mathcal{B} = \left\{\mathcal{A}\cup\mathcal{B}\right\}''$,
and $'$ denotes the commutant, as usual.  Composition
of weights or automorphisms is denoted by $\circ$, 
so that, for example, 
$\tau\circ\theta_s(\msf{a}) = \tau(\theta_s(\msf{a}))$.  
The star operation on operators will be denoted by $*$; 
when these operators act on a Hilbert space, this operation 
agrees with the Hermitian adjoint, commonly denoted 
as $\dagger$ in other works.  All weights and operator-valued 
weights considered in the work are assumed to be 
{\it normal}, which is a statement of continuity 
with respect to the $\sigma$-weak topology
(see \cite[Definition VII.1.1, Definition IX.4.12]{TakesakiII}).

\section{Quantization of the rolling scalar}
\label{sec:inflalg}

The focus of this work will be on the slow-roll inflation
example considered by Chen and Penington \cite{Chen2024}.  They 
considered a scalar field in 4-dimensional de Sitter space in
the semiclassical gravity limit with $G\rightarrow 0$.  
The dynamical degrees of freedom in their model are the 
modes of the scalar field and the free gravitons.  The 
small $G$ limit suppresses back-reaction on the metric, leading
to an exactly de Sitter background spacetime on which 
the fields are quantized.  

The algebra of interest is the collection of modes in causal contact 
with a single worldline in the spacetime, which heuristically
can be viewed as the algebra accessible to an observer in the 
spacetime.   These degrees of freedom coincide with 
quantum fields localized within a single static patch of 
de Sitter space.  The gauge-invariant algebra is then
constructed from these degrees of freedom as the set of 
operators invariant under the static-patch time translation,
which is the isometry of de Sitter space that is future
directed in the patch, past-directed in the complementary patch,
and acts at a boost at the bifurcation surface on the boundary.  
We will refer to this as the boost isometry.  

For standard fields in de Sitter space, one expects this 
boost to act ergodically on the operators localized to the 
static patch, leading to the conclusion that the gauge-invariant
algebra is trivial.  This point was the original motivation
for CLPW \cite{Chandrasekaran2022a} 
to introduce an explicit gravitating observer 
in order to obtain a nontrivial algebra for the de Sitter 
static patch.  Doing so allowed for the construction 
of relational observables dressed to the observer's clock, 
resulting in operators that are invariant under the static patch boost
symmetry.  Chen and Penington proposed an alternative 
resolution that does not require the introduction 
of an explicit observer into the theory.  Instead, they
chose a potential for the scalar field such that 
it perpetually rolls to lower values, so that there is no normalizable
stationary state for the field.  The scalar then provides a notion
of clock to which the other degrees of freedom can be dressed,
and Chen and Penington argue that this leads to nontrivial
operators invariant under the boost flow.  

In the present work, we will analyze a simplified version
of Chen and Penington's model that maintains the important
features of the construction.  The simplification
is to work in $2D$ de Sitter space.  Doing 
so eliminates the need to consider gravitons as well 
as considerations of higher spherical harmonic modes 
for the scalar field.  This model is still
relevant because the interesting effect leading to the 
type $\ttwo_\infty$ algebra in Chen and Penington's work
happens in the $s$-wave scalar sector, and one can argue
that the inclusion of other fields such as gravitons 
or higher angular momentum modes does not change the 
essential conclusions.  In particular, the arguments for 
nontrivial, boost-invariant operators in the $2D$ model
are exactly the same as in the $4D$ case, as is the argument
that the resulting algebra is a crossed product.  The $2D$ model
has the advantage that explicit computations are somewhat
easier to perform, so we will focus on this case, and comment
in  various places on how the argument would generalize 
to higher dimensions.  

The metric of $\text{dS}_2$ in conformally compactified
coordinates is 
\beq
ds^2 = \frac{\ell^2}{\cos^2(T)}(-dT^2 + d\chi^2),
\eeq
with $\chi$ a $2\pi$-periodic coordinate on the spatial
circle, and $T\in(-\frac{\pi}{2},\frac{\pi}{2})$.  The scalar field is taken to be massless with an 
exactly linear potential, whose action is 
\begin{align}
S &= \int d^2x \sqrt{-g}\left(-\frac12\nabla_a \phi \nabla^a\phi
- c\phi\right) \label{eqn:scalact}\\
&= \int d T \int_0^{2\pi} d\chi  \left(
\frac12\dot{\phi}^2 
- \frac12\phi'^2 - \frac{c\ell^2}{\cos^2(T)}\phi\right ).
\label{eqn:Scoord}
\end{align}
The linear potential is a good approximation for an inflaton
field in a slow-roll regime far from the minimum of the potential.
In this model, the linear potential pushes the scalar to smaller
values under time evolution, so that it never settles down to a 
stationary configuration.  

Since this is a free field theory, it can be quantized by imposing canonical commutation 
relations in the time slicing defined by the coordinate $T$.
The momentum field on such a slice is given by
\beq
\pi(\chi) =  \dot\phi(\chi),
\eeq
and the equal-time commutation relation reads
\beq \label{eqn:ccr}
[\phi(\chi_1), \pi(\chi_2)] = i \delta(\chi_1-\chi_2).
\eeq
To construct the Hilbert space, we also need 
to specify an appropriate vacuum state on which to build the 
representation of the commutation relation.  Here, there is an
important subtlety related to the existence of a zero mode
for the massless scalar.  
The action (\ref{eqn:scalact}) is invariant up to a $\phi$-independent
constant under the shift transformation $\phi(x)\rightarrow
\phi(x) + a$, necessitating a separate treatment of the spatially
homogeneous modes of the scalar field.  The zero modes  are
responsible for the well-known issues with IR divergences
in the massless scalar 2-point function in de Sitter space
\cite{Allen:1985ux}.  Our approach to handling these zero modes will be 
as in 
\cite{Chen2024} (see also \cite{Kirsten:1993ug, Tolley:2001gg, Kudler-Flam:2025pol}): the 
zero mode sector will consist of a single canonical pair
quantized in the standard way on Hilbert space
$\hs_0 = L^2(\mathbb{R})$.  The main subtlety is that there
will not be a preferred normalizable de-Sitter-invariant 
state for this 
zero mode sector, and hence all normalizable states must 
involve a specification of a wavefunction for the zero
modes.  Nevertheless, there is a distinguished unnormalizable
state, the Bunch-Davies weight, which will play a crucial 
role in the subsequent discussion.  

A second point to note is that because the potential in 
(\ref{eqn:scalact}) is linear in $\phi$, it can be 
eliminated from the action, up to a total derivative, by
a field redefinition.  This means we can write
\beq \label{eqn:phicl}
\phi = \phi_\text{cl} + \varphi,
\eeq
where $\phi_\text{cl}$ is a $c$-number solution to the classical
equations of motion, and $\varphi$ is the quantum
operator representing the 
perturbation around this background. $\phi_\text{cl}$ 
is straightforward to determine: the equations of 
motion in this coordinate system is an inhomogeneous 
wave equation
\beq
\ddot{\phi} - \phi'' = -\frac{c\ell^2}{\cos^2(T)}.
\eeq
Taking a spatially homogeneous profile $\phi' = 0$, this equation
can  be integrated to obtain 
\beq
\phi_\text{cl} = c\ell^2 \log(\cos(T)).
\eeq

The field operator $\varphi$ then satisfies the homogeneous 
wave equation
\beq
\ddot{\varphi} - \varphi'' = 0.
\eeq
Quantization now proceeds as usual.  The zero mode 
solutions have $\varphi' = 0$ and are given by $\varphi(T)
= \frac{1}{\sqrt{2\pi}}(\varphi_0 + \pi_0 T)$ for 
real coefficients $\varphi_0, \pi_0$.  The remaining solutions
have spatial profiles $e^{i n\chi}$ for $n\in(\mathbb{Z}-\{0\})$
and oscillate in time with frequencies $\omega = \pm|n|$.  
Hence the field $\varphi$ admits the expansion\footnote{The 
solution $f_n(T,\chi)$ 
multiplying the coefficient $a_n$
is chosen to have unit Klein-Gordon inner product, 
 $i\int_0^{2\pi} d\chi (f_n^*\dot{f}_n - f_n \dot{f}_n^*) = 1$.
} 
\beq
\varphi = \frac{1}{\sqrt{2\pi}}\varphi_0 +\frac{T}{\sqrt{2\pi}}
\pi_0 +
\sum_{n\in(\mathbb{Z}-\{0\})}\frac{e^{in\chi}}{\sqrt{4\pi|n|}}
\left(e^{-i|n|T} a_n + e^{i|n|T} a_{-n}^*\right)
\eeq

The vacuum we will use to complete the quantization
is the Bunch-Davies weight.  The term ``weight'' refers to the 
fact that this state will not be normalizable due to 
its behavior in the zero-mode sector.  Nevertheless, it can 
be used to construct the Hilbert space for the field quantization
using a GNS-like procedure known as 
the semicyclic representation \cite[Section VII.1]{TakesakiII}.  This weight is the unique de-Sitter-invariant
weight with the appropriate short-distance behavior for the 
$\phi(x)$ two-point function.  It can be computed by evaluating
the classical Euclidean action on solutions that are 
regular at the south pole of the 2-sphere, which is the 
Euclidean continuation of $\text{dS}_2$. This south
pole is located at $T = +i\infty$, and so the constant solution
$\varphi_0$ as well as the oscillating
solutions involving $a_n^* e^{i|n|T}$ are the appropriate 
ones.  

The general solution with this boundary
condition can be written
\beq
\phi(T,\chi) = \left(\phi_\text{cl}(T)+ic\ell^2 T\right)+\frac{1}{\sqrt{2\pi}}
\sum_{n\in\mathbb{Z}}\varphi_n e^{in\chi} e^{i|n|T},
\eeq
where $\varphi_n$ are the Fourier coefficients of the field 
value at the $T=0$ surface, and $\phi_\text{cl}(T)+ ic\ell^2 T$
is the background solution that is regular at $T\rightarrow +i\infty$.  The action (\ref{eqn:Scoord})
on this solution evaluates to 
\begin{align}
iS = i\int dT \left[-\frac{\sqrt{2\pi}c\ell^2}{\cos^2(T)} \varphi_0-
2\sum_{n>0}\varphi_n\varphi_{-n} n^2 e^{2i|n|T} \right] +
\text{const.},
\end{align}
where the constant only involves contributions from 
the background solution $\phi_\text{cl} + ic\ell^2 T$.  
To arrive at the final result, we perform the $T$ integral
along a contour that follows the imaginary axis from $+i\infty$
to $0$.  This leads to the expression for the Bunch-Davies 
wavefunction
\beq \label{eqn:psiBD}
\Psi_\text{BD} \propto \exp[i S] = 
\exp\left[-\sqrt{2\pi} c\ell^2 \varphi_0 - 
\sum_{n>0}n \varphi_n\varphi_{-n}\right].
\eeq

The first point to note about this wavefunction is that 
it is not normalizable.  It depends exponentially on the 
zero mode $\varphi_0$, which ranges from $-\infty$ to $\infty$, 
hence the contribution to the norm coming from 
the zero mode sector is divergent.  This is the manifestation
of the usual IR divergence for massless fields in 
de Sitter, and indeed we would find that the 2-point function
$\langle\Psi_\text{BD}|\phi(x) \phi(y)|\Psi_\text{BD}\rangle$ 
is divergent in this state.  This same divergence is present
for the standard massless scalar potential $V(\phi) = 0$ obtained
by setting $c=0$, since the constant wavefunction for $\varphi_0$ 
is still a nonnormalizable state.  However, we can still
use $\Psi_\text{BD}$ to construct a Hilbert space for the 
scalar field, as long as we interpret $\Psi_\text{BD}$ as a 
semifinite weight, meaning it assigns finite expectation 
values only to a dense subset of operators in the theory.  Such operators must involve functions of $\varphi_0$ that decay
rapidly as $\varphi_0\rightarrow -\infty$ to cancel
the exponential growth in the wavefunction.  The normalizable 
states of the theory must then all involve a nontrivial 
wavefunction of $\varphi_0$ that provides an effective IR 
cutoff on correlation functions; see, for example,
the vacua constructed in 
\cite{Allen:1987tz, Kirsten:1993ug}.  
Since all such wavefunctions
will break some de-Sitter symmetries, we see in this 
case that there are no normalizable dS-invariant states.  

The second point to note is that the representation we 
obtain of the field algebra using the vacuum
weight $\Psi_\text{BD}$ is isomorphic to the field 
algebra of the standard massless scalar with $c=0$.  Both field
algebras are labeled by the modes $\varphi_n$ and their conjugate
momenta, and the only difference between the Bunch-Davies 
weights for $c=0$ and $c\neq 0$ is the wavefunction
for the zero mode.  Since the zero mode sector 
represents only a single degree of freedom,
and the canonical commutation relations have a unique 
representation in finite dimensions, we see that the 
quantizations with respect to $\Psi_\text{BD}(c=0)$
and $\Psi_\text{BD}(c\neq 0)$ must be unitarily 
equivalent.  

The important difference between the two quantizations 
is how the de Sitter isometries act on the field modes
$\varphi_n$.  For any Killing vector $\xi^a$ of $\text{dS}_2$,
the generators $H_\xi$ are defined to act on the field 
$\phi(x)$ via the Lie derivative,
\beq
[H_\xi, \phi(x)] = -i\lie_\xi\phi(x) = -i \xi^a\nabla_a\phi(x).
\eeq
For the standard massless theory, the background solution 
satisfies $\phi_\text{cl} = 0$, and so by (\ref{eqn:phicl}),
the perturbation $\varphi$ and the original field $\phi$ are 
equal.  On the other hand, when $c\neq 0$, the background 
solution is nonzero, and so $\varphi$ and $\phi$ differ
by the $c$-number function $\phi_\text{cl}(T)$.  
Since $H_\xi$ commutes with all $c$-numbers, we find that its 
action on the field perturbation is modified,
\beq \label{eqn:Hximod}
[H_\xi, \varphi(x)] = [H_\xi, \phi(x)] = -i\lie_\xi \varphi(x)
- i \lie_\xi \phi_\text{cl}(T).
\eeq
This modified action of $H_\xi$ on $\varphi(x)$ ends up being 
responsible for the existence of nontrivial
operators in the static patch that are invariant under 
the boost isometry.  

The modified action can also be explained from the 
form of the stress tensor in the slow-roll theory.  The scalar
stress tensor is given by 
\beq
T_{ab} = \nabla_a\phi \nabla_b \phi - \frac12 g_{ab}\nabla_c\phi
\nabla^c \phi  - c g_{ab} \phi.
\eeq
From this, we can construct the generator of the 
static patch boost by smearing against the Killing vector
$\xi^a = \cos\chi \cos T\partial_T^a  - \sin\chi \sin T
\partial_\chi^a$, and integrating over a Cauchy slice.  
Taking the $T=0$ slice, this gives
\beq \label{eqn:Hxi}
H_\xi = \int d\chi \xi^T T_{TT} = \int d\chi \cos\chi\left(
\frac12\pi(\chi)^2 + \frac12\phi'(\chi)^2 + c\ell^2\phi(\chi)\right).
\eeq
Noting that $\phi_\text{cl}$ and $\dot\phi_\text{cl}$ 
vanish on the $T=0$ slice, we see that 
the term linear in $\phi(\chi)$ accounts for the modified
action of $H_\xi$ on the field operators $\varphi$. 

Finally, it is important to emphasize that $|\Psi_\text{BD}\rangle$ 
defines a KMS weight for operators localized in the 
static patch, defined as the region $|T| + |\chi|<\frac{\pi}{2}$.
There are plenty of operators smeared only in the static patch
that have a nontrivial component in the zero mode 
sector: take, for example, the spatially smeared field operator
\beq
\phi_f = \int d\chi f(\chi) \phi(\chi),
\eeq
with $f(\chi)$ supported in $\chi\in[-\frac{\pi}{2},\frac{\pi}{2}]$,
and $\int d\chi f(\chi) \neq 0$.  Taking bounded functions of this 
operator that decay rapidly enough at large negative arguments 
will result in an operator with a finite expectation
value in $|\Psi_\text{BD}\rangle$. We expect the full
collection of operators with finite expectation
values to form a weakly dense subalgebra of $\alg$,
the full von Neumann algebra of operators localized
to the static patch.  This implies that $|\Psi_\text{BD}\rangle$
defines a semifinite weight on $\alg$.  The boost isometry
generated by $H_\xi$ which fixes the static patch 
leaves $|\Psi_\text{BD}\rangle$ invariant, and one can 
argue from the Euclidean path integral construction
that correlation functions satisfy a KMS condition
for this flow \cite{Chen2024},
\beq\label{eqn:BDKMS}
\langle\Psi_\text{BD}|\msf{a}_{t} \msf{b}|\Psi_\text{BD}\rangle
=\langle\Psi_\text{BD}|\msf{b} \msf{a}_{t+2\pi i}
|\Psi_\text{BD}\rangle,
\eeq
for operators $\msf{a}$ and $\msf{b}$ with finite
expectation value in $|\Psi_\text{BD}\rangle$, and 
where $\msf{a}_t \defeq e^{iH_\xi t}\msf{a}e^{-iH_\xi t}$.
This implies that $H_\xi$ must generate the modular 
automorphism of $\alg$ for the weight 
corresponding to $|\Psi_\text{BD}\rangle$
\cite[Theorem VIII.1.2]{TakesakiII}, with the modular
Hamiltonian $h$ given by 
\beq
h = 2\pi H_\xi.
\eeq

\section{Integrable Bunch-Davies weight}
\label{sec:intweight}

The inflationary gravitational algebra $\agrav$ is 
obtained 
as the boost-invariant subalgebra of $\alg$, the algebra
of operators localized to the static patch.
As discussed above, the boost generates an
automorphism of $\alg$ which 
we denote by $\sigma_t(\msf{a}) =e^{i2\pi H_\xi t}\msf{a}e^{-i2\pi H_\xi t}$.
One way to try to form boost-invariant operators 
is to average a non-invariant operator $\msf{a}$ over time,
\beq \label{eqn:Tovw}
\Tovw(\msf{a}) = \int_{-\infty}^\infty dt\, \sigma_t(\msf{a}).
\eeq
Assuming this integral converges, the image of $\Tovw$ will always
result in an operator that commutes with $H_\xi$.  In the 
present context, we expect this integral
to converge on a dense set of operators in the 
static patch.  The reason comes from the modified 
commutation relation (\ref{eqn:Hximod}) for the action of $H_\xi$
on the field operators.  This equation 
shows that in addition to their standard time evolution in the 
patch, the field operators pick up a $c$-number shift coming
from the fact that the background solution is not invariant
under the Killing flow.  Hence by forming bounded combinations
of the field operators that decay at large arguments, such as 
$\exp\left[-\left(\int f \phi\right)^2\right]$ where $f$ is 
a smearing function supported in an open region in the 
static patch, one can construct operators for which the 
time average (\ref{eqn:Tovw}) converges, and we expect
the full set of such operators is weakly dense in $\alg$.  
When this occurs, 
the automorphism $\sigma_t$ is called {\it integrable}.

The time average operation $\Tovw$ is an example of a semifinite
{\it operator-valued weight}
\cite{Haagerup1979I, Haagerup1979II}\cite[Section IX.4]{TakesakiII}, and since these will be used throughout this 
section, we take a moment here to review their essential
properties.  Operator-valued weights are unbounded versions
of conditional expectations, much like how weights are 
unbounded versions of states.  In general, if $\agrav\subset
\alg$ is a von Neumann subaglebra, an operator-valued weight 
is defined as a linear map map from a definition subaglebra 
$\mathfrak{m}_{\Tovw} \subset \alg$ to $\agrav$ satisfying the 
bimodule property
\beq
\Tovw(\tilde{\msf{a}} \msf{b} \tilde{\msf{c}}) = 
\tilde{\msf{a}} \Tovw(\msf b) \tilde{\msf{c}}, \qquad \tilde{\msf{a}},
\tilde{\msf{c}} \in \agrav, \; \msf{b} \in \mathfrak{m}_{\Tovw}.
\eeq
$\Tovw$ assigns an infinite value to 
any operator not contained in the definition 
subalgebra.  When $\mathfrak{m}_{\Tovw}$ is 
$\sigma$-weakly dense in $\alg$, $\Tovw$ is said to be {\it semifinite}.  
Unlike conditional expectations, operator-valued weights 
 are {\it not} required to be idempotent, i.e.\ in general 
$\Tovw\circ \Tovw \neq \Tovw$.  In fact, for the operator-valued weights of 
interest in the present work, none of the operators in 
$\agrav$ are contained in the definition domain, so that 
$\Tovw\circ\Tovw = \infty$.  It can be helpful to view $\Tovw$ 
as an unnormalized conditional expectation
satisfying $\Tovw\circ \Tovw = N \, \Tovw$, where the 
normalization coefficient $N$ can be infinite.

In addition to being integrable, the automorphism $\sigma_t$ 
also satisfies a KMS condition (\ref{eqn:BDKMS}) 
for the Bunch-Davies weight
$\omega_\text{BD} = 
\langle\Psi_\text{BD}|\cdot|\Psi_\text{BD}\rangle$.  
The KMS condition for $\sigma_t$ immediately implies that 
$\sigma_t$ generates modular flow for the weight 
$\omega_\text{BD}$ 
\cite[Theorem VIII.1.2]{TakesakiII}, in which case the algebra $\agrav$ consists
of operators invariant under modular flow, and is called 
the {\it centralizer of the weight} $\omega_{\text{BD}}$.  
The weight $\omega_\text{BD}$ is called an {\it integrable
weight} since its associated modular flow is an 
integrable automorphism.  

Centralizers of weights are interesting because they often 
are semifinite,
meaning they are endowed with a semifinite trace.  
This is easy to see for a normalized state $\varphi$
which assigns a finite value to every 
operator in $\alg$.  Operators in the centralizer $\alg_\varphi$
are invariant under modular flow, which for type $\tone$ or $\ttwo$
algebras would mean these operators commute with the 
density matrix $\rho_\varphi$ for the state $\varphi$.  But this 
means $\varphi(\msf{a}\msf{b}) = \varphi(\msf{b}\msf{a})$ for $\msf{a}, \msf{b}\in
\alg_\varphi$, so $\varphi$ defines a trace on 
$\alg_\varphi$.  Even in the type $\tthr$ case where density matrices 
are ill-defined, operators in the centralizer still 
satisfy the tracial property $\varphi(\msf{ab})  = \varphi(\msf{ba})$
\cite[Theorem VIII.2.6]{TakesakiII}.
The argument is more subtle when $\varphi$ is a weight,
since it may not be semifinite when restricted to $\alg_\varphi$, meaning
it will not assign a finite value to a dense subset of operators in 
$\alg_\varphi$.  A weight that remains semifinite when restricted to 
its centralizer is called {\it strictly semifinite}.  For such weights,
there is a conditional expectation $\cal{E}: \alg\rightarrow \alg_\varphi$
that preserves $\varphi$.

On the other hand, integrable weights are never strictly semifinite,
since one can straightforwardly show that an integrable weight  assigns
an infinite value to any element 
of the centralizer.  This just follows from
the fact that an integrable weight $\omega$ can be written as $\omega = \tau_\omega
\circ \Tovw$, where $\Tovw$ is the operator-valued weight (\ref{eqn:Tovw}), and 
$\tau_\omega$ is a semifinite 
weight on the centralizer.  Whenever $\msf{a}\in
\alg_\omega$, it is invariant under the flow $\sigma_t$,
and so the integral in (\ref{eqn:Tovw}) clearly diverges,
showing that $\omega$ is not semifinite on its centralizer.  However,
Haagerup's theorem applied to the operator-valued weight $\Tovw$ implies that 
$\tau_\omega$ must induce a modular flow on $\alg_\omega$
that agrees with that of $\omega$
\cite{Haagerup1979II}\cite[Theorem IX.4.18]{TakesakiII}.  But since $\alg_\omega$ is fixed by 
modular flow, it must be that modular flow of $\tau_\omega$ is trivial,
and hence $\tau_\omega$ 
defines a trace on $\alg_\omega$.  Thus we see there 
are two interesting 
cases where the centralizer is semifinite: either when the weight is 
strictly semifinite and there is a conditional expectation, or when
the weight is integrable and there is an operator-valued weight.
In the most general case, the centralizer of a weight $\varphi$ is 
semifinite if and only if there exists a semifinite 
operator-valued weight from $\alg$ to the centralizer $\alg_\varphi$
\cite[Theorem 5.7]{Haagerup1979II}.
Note there are examples of weights $\varphi$ for which the centralizer
is type $\tthr$, in which case there is no trace defined on the 
centralizer and correspondingly no operator-valued weight that preserves
the weight $\varphi$
\cite{Haagerup1977}.

There is a detailed theory of integrable weights on type $\tthr$ 
von Neumann algebras  developed by Connes and Takesaki,
which is broadly referred to as
the {\it flow of weights} \cite{Connes1977}
\cite[Section XII.4]{TakesakiII}.  Integrable weights are classified
(up to equivalence---see section \ref{sec:vacds}
for details of this comparison theory for weights) 
by weights on an abelian algebra $\mc{C}_{\alg}$ that
appears as the center of the modular crossed product
algebra $\alg\rtimes_\sigma\mathbb{R}$.  This classification is particularly
simple in the type $\tthr_1$ case, since the crossed product 
algebra is a factor, and hence $\mc{C}_{\mc{A}} = \mathbb{C}\mathbbm{1}$.  
The weight $\tilde\omega_{\mc{C}}$ on $\mc{C}_{\mc{A}}$ associated to 
an integrable weight $\omega$ is then just determined by 
a positive number $w=\tilde\omega_{\mc C}(\mathbbm{1})
\in (0,\infty]$. 
We can easily ascertain the value of this 
number by first noting that 
$\mc{C}_{\alg}$ can be identified 
with the (trivial) center of the centralizer, 
$Z(\alg_\omega)$.  We then define a weight $\tau_\omega$ on $\alg_\omega$
by demanding that $\omega = \tau_\omega \circ \Tovw$, with $\Tovw$ the 
operator-valued weight (\ref{eqn:Tovw}).  Notably, this unambiguously
fixes the normalization of the weight $\tau_\omega$, and its value 
on the identity determines $w = \tau_\omega(\mathbbm{1})$
\cite[Corollary II.3.2]{Connes1977}.  When
$w<\infty$, we can define a normalized trace $\hat\tau_\omega = 
\frac{1}{w}\tau_\omega$ on $\alg_\omega$, showing in this case 
that $\alg_\omega$ is type $\ttwo_1$.  When $w=\infty$, the weight $\omega$
is known as a {\it dominant weight}, and the centralizer $\alg_\omega$ 
is a type $\ttwo_\infty$ algebra isomorphic to the modular 
crossed product.  As we discuss in section \ref{sec:vacds}, the integrable
weights with $w<\infty$ 
are relevant to constructions of gravitational algebras involving 
an observer in the de Sitter static patch.  On the other hand, the 
slow-roll inflation example results in a type $\ttwo_\infty$ 
algebra, which can be confirmed due to the existence 
of an automorphism $\theta_s$ of $\alg_\omega$ that rescales 
the trace according to $\tau_\omega \circ \theta_s = e^{-s} \tau_\omega$.  
Such an automorphism is only possible for type $\ttwo_\infty$ algebras,
and hence
implies that the Bunch-Davies weight
is dominant.   As explained in more detail
in section \ref{sec:crossprod}, this automorphism
comes from the shift symmetry of the massless scalar field.  

Once we have identified $\omega_\text{BD}$ as a dominant weight, we are 
able to conclude a number of properties of its centralizer,
the gravitational algebra $\agrav$.  
As stated above, it immediately follows that the centralizer
is a type $\ttwo_\infty$ {\it factor}, isomorphic to the modular
crossed product algebra.  This verifies the expectation
from Chen-Penington that the gravitational algebra has a trivial
center.  
In particular, since $\alg$ is the 
unique hyperfinite $\tthr_1$ factor $\rinj_\infty$, its centralizer
must be the unique hyperfinite $\ttwo_\infty$ factor $\rinj_{0,1}$,
invoking, e.g., the result that the fixed point algebra of an injective
algebra with respect to the action of a locally compact amenable 
group must be injective (see \cite[Theorem XV.3.16]{TakesakiIII}).  
A second property relates to a renormalization that occurs between
the Bunch-Davies weight $\omega_\text{BD}$ and the trace $\tau_\text{BD}$
on $\agrav$.  As we have seen, $\omega_\text{BD}$ diverges
on any element of the centralizer $\agrav$, but this divergence 
is associated with a universal infinite rescaling that comes 
from passing the operator through the operator-valued weight $\Tovw$.  
On elements of $\agrav$, we can formally write $\Tovw(\tilde{\msf a})
= N \tilde{\msf{a}}$, viewing $N$ as a divergent constant.  Then
since $\omega_\text{BD} = \tau_{\text{BD}}\circ \Tovw$, we find that 
\beq
\tau_\text{BD}(\tilde{\msf a}) = \frac{1}{N}\tau_\text{BD}(
\Tovw(\tilde{\msf a})) = \frac{1}{N} \omega_\text{BD}(\tilde{\msf a}),
\eeq
showing that the trace $\tau_\text{BD}$ can formally be viewed as an
infinitely rescaled version of $\omega_\text{BD}$ on elements of 
$\agrav$.  The unbounded operator-valued weight $\Tovw$ thus provides 
a mathematically rigorous characterization of this 
formally infinite renormalization of $\omega_\text{BD}$
A related discussion of this infinite renormalization
appears in \cite{Chen2024}, and we see here it has a natural
explanation in terms of the operator-valued weight $\Tovw$. 

An important comment can be made at this point justifying
the choice to focus on two dimensions.  The reason
we do this is that the arguments leading to the 
conclusion that $\omega_\text{BD}$ is 
dominant are unaffected by the inclusion
of higher angular momentum modes.  For
example, the four-dimensional Bunch-Davies weight 
computed in \cite{Chen2024} factorizes between
the spherically symmetric sector and the rest of the algebra.
Hence in that case one can write the quantum field
algebra as $\alg = \alg_{m=0}\otimes \alg_{m>0}$, with
$\alg_{m=0}$ the spherically symmetric algebra.  The
Bunch-Davies weight then factorizes as $\omega_\text{BD} =
\omega_0 \otimes\omega_{>0}$.  The spherically symmetric
weight $\omega_0$ behaves much like the Bunch-Davies 
weight for the two-dimensional model considered in 
the present work.  In particular, $\omega_0$ is a dominant
weight on $\alg_{m=0}$.  This immediately implies that 
the higher dimensional Bunch-Davies weight is dominant,
since any time $\omega_D$ is a dominant weight on an
algebra $\alg$, the product weight $\omega_D\otimes \psi$
is dominant on the product algebra $\alg\otimes\mathcal{B}$,
where $\psi$ is any faithful weight on 
$\mathcal{B}$.\footnote{This follows easily from the property
that when $\omega_D$ is dominant, there exists a unitary
$\msf{u}_\lambda \in\alg$ for any $\lambda>0$ such that 
$\omega_D(\msf{u}^*_\lambda \cdot \msf{u}_\lambda)
= \lambda \omega_D(\cdot)$.  This same unitary rescales 
the factorized weight $\omega_D\otimes \psi$, and 
since being unitarily equivalent to the rescaled
weight is the defining property of a dominant
weight \cite[Theorem II.1.1]{Connes1977}\cite[Theorem XII.4.18]{TakesakiII}, 
we see that $\omega_D\otimes \psi$ is 
dominant.  
}
The same argument explains why one does not need to consider
the graviton contribution in detail in four or 
higher dimensions.  The gravitons simply appear as an additional
factor in the algebra, and since the scalar Bunch-Davies weight
is already dominant, tensoring in the graviton 
contribution also produces a dominant weight.  Hence 
including gravitons or other matter fields does not 
affect the conclusion that $\omega_\text{BD}$ is dominant,
and therefore still has a type $\ttwo_\infty$ centralizer.

\section{Crossed product description of gravitational algebra}
\label{sec:crossprod}

Although the identification of $\agrav$ as the centralizer of a 
dominant weight on the type $\tthr_1$ algebra $\alg$ immediately
implies that it is isomorphic to a modular crossed product, 
the isomorphism is not canonical at this point.  
In order to
to canonically identify $\agrav$ as a crossed product 
algebra, it is necessary to decompose $\agrav$ into a type $\tthr_1$ 
subalgebra and a collection of operators generating the action
of a modular automorphism on the subalgebra.  
Arriving at this canonical decomposition
requires the second key feature of the inflationary
gravitational algebra, which is the existence of a preferred 
family of trace-scaling automorphisms $\theta_s$.  
These automorphisms correspond with the
shift symmetry of the massless scalar field.  
The operators invariant under the shift symmetry 
will form the type $\tthr_1$ subalgebra, and $\agrav$ is then
represented as a crossed product by adding an additional
operator that plays the role of an observer Hamiltonian.  
There ends up being a further ambiguity in determining 
this observer Hamiltonian, which can be interpreted as 
a freedom to choose a quantum reference frame for the description
of the algebra $\agrav$.  

The generator of the shift symmetry for the scalar field
 is the constant momentum mode, which we write as 
\beq \label{eqn:Pi}
\Pi = \frac{-1}{2\sqrt{2\pi} c\ell^2} \pi_0 = \frac{-1}{4\pi c\ell^2}
\int_0^{2\pi} d\chi \pi(\chi).
\eeq
Since it generates a shift in the scalar field $\phi(x)$, it acts 
only on the zero mode $\varphi_0 = \frac{1}{\sqrt{2\pi}}\int_0^{2\pi} 
d\chi \phi(\chi)$, shifting it by an operator proportional
to the identity,
\beq
e^{is\Pi} \varphi_0 e^{-is\Pi} = \varphi_0 -\frac{s}{2\sqrt{2\pi} c\ell^2}
\mathbbm{1}.
\eeq
From this relation, we see that the shift symmetry rescales the Bunch-Davies
weight (\ref{eqn:psiBD}),
\beq
e^{-is\Pi} \Psi_\text{BD} = e^{-\frac{s}{2}} \Psi_{\text{BD}}.
\eeq
It is also straightforward to verify using
the canonical commutation relations (\ref{eqn:ccr}) that $\Pi$ commutes 
with the boost Hamiltonian (\ref{eqn:Hxi}), which means that $\Pi$ 
also generates an automorphism of the centralizer $\agrav$.  Denoting
the automorphism as $\theta_s = \Ad(e^{is\Pi})$, we see 
that it must commute with the time-average operator-valued weight
$\Tovw$, and hence must rescale the trace on $\agrav$:
\beq
(\tau\circ \theta_s) \circ \Tovw = \tau\circ \Tovw\circ \theta_s
=\omega_\text{BD}\circ\theta_s = e^{-s}\omega_\text{BD} = (e^{-s}\tau)
\circ \Tovw,
\eeq
from which we conclude
\beq
\tau\circ\theta_s = e^{-s} \tau.
\eeq

The factor $e^{-s}$ by which the trace is rescaled is known as the 
{\it module} of the automorphism
$\theta_s$ when acting on $\agrav$, and its 
exponential dependence on the parameter $s$ follows from
the requirement that $\theta_s$ be a homomorphism from $\mathbb{R}$
into $\aut(\agrav)$, the automorphism group of $\agrav$.  This
implies that $\theta_{s+u} = \theta_s \circ \theta_u$, 
which then imposes that the modules  satisfy
$\Mod(\theta_{s+u}) = \Mod(\theta_s)\Mod(\theta_u)$.
The only solution to this relation is $\Mod(\theta_s) = e^{-\alpha s}$,
and by rescaling the flow parameter $s$ 
we can set $\alpha = 1$ whenever the module is not identically
equal to $1$.  

One-parameter groups of trace-scaling automorphisms on 
semifinite von Neumann algebras have been classified by 
Takesaki in developing the structure theorems for 
type $\tthr$ von Neumann algebras \cite{Takesaki1973}\cite[Section
XII.1]{TakesakiII}.  We are particularly interested in the structure
of the fixed-point algebra $\agravss = \agrav^\theta$, consisting 
of all operators invariant under the automorphism $\theta_s$.  
As we discuss below, this fixed point algebra must be a type $\tthr_1$
factor
when $\agrav$ is a type $\ttwo_\infty$
factor.  An essential point in reaching this conclusion is the 
observation that every one parameter group
of trace-scaling automorphisms is 
integrable 
(see the proof of Lemma XII.1.2 of \cite{TakesakiII}
or the proof of Theorem III.5.1(ii) of \cite{Connes1977}),
so that the integral
\beq
\Sovw(\tilde{\msf{a}}) = \int_{-\infty}^\infty ds\, \theta_s(\tilde{\msf{a}})
\eeq
defines a semifinite operator-valued weight $\Sovw$ from $\agrav$ to 
$\agravss$.  
This allows any weight $\omega$ on $\agravss$ to be lifted to a weight
$\tilde\omega$ on $\agrav$ by composing it with the operator-valued 
weight, $\tilde\omega = \omega\circ \Sovw$.  
Because $\agrav$ possesses a trace, the weight $\tilde\omega$ 
can be represented by a positive, self-adjoint 
density matrix $\rho_\omega$ affiliated with $\agrav$ 
according to the 
relation
\beq
\tilde\omega(\tilde{\msf{a}}) = \tau(\rho_\omega \,\tilde{\msf{a}})
\eeq
The density matrix $\rho_\omega$ is an eigenoperator of the automorphism
$\theta_s$, which follows from the fact that the weight $\tilde\omega$
is invariant under $\theta_s$.  This implies that 
\beq
\tau(\theta_s(\rho_\omega) \tilde{\msf{a}})
= e^{-s}\tau(\rho_\omega \theta_{-s}(\tilde{\msf{a}}))
=e^{-s}\tilde\omega(\theta_{-s}(\tilde{\msf{a}}))
=e^{-s}\tilde\omega(\tilde{\msf{a}})
=e^{-s}\tau(\rho_\omega \tilde{\msf{a}}),
\eeq
and thus
$
\theta_s(\rho_\omega) = e^{-s} \rho_\omega.
$
Equivalently, it implies that $\hobs_\omega = \log\rho_\omega$ transforms by shifts under the 
automorphism,
\beq
\theta_s(\hobs_\omega) = \hobs_\omega - s.
\eeq

The operator $\hobs_\omega$ plays a role  analogous  to the observer 
Hamiltonian in the CLPW gravitational algebra in vacuum de Sitter
or the asymptotic charges in the gravitational crossed product 
constructions 
\cite{Witten2021, Chandrasekaran2022a, Chandrasekaran2022b,
Jensen2023, Kudler-Flam:2023qfl, Faulkner:2024gst}.  Crucially, it 
generates an automorphism $\alpha_t$ of the subalgebra $\agravss$
by conjugation $\alpha_t(\msf{a}_0) = e^{-it\hobs_\omega}
\msf{a}_0 e^{it\hobs_\omega}$, which
follows from the fact that
\beq
\theta_s\left(e^{-it\hobs_\omega} \msf{a}_0 e^{it\hobs_\omega}\right)
= e^{ist} e^{-it\hobs_\omega} \theta_s(\msf{a}_0) e^{it\hobs_\omega}
e^{-ist}
= e^{-it\hobs_{\omega}} \msf{a}_0 e^{it\hobs_\omega},
\eeq
showing that $\alpha_t(\msf{a}_0) \in
\agravss$ since it is invariant under $\theta_s$.  This automorphism
is just the modular automorphism
for the weight $\omega$, since $\rho_\omega$ generates modular
flow on $\agrav$ of a weight that is fixed by $\Sovw$ 
\cite{Haagerup1979I}\cite[Theorem IX.4.18]{TakesakiII}.  
In fact, the existence of the unitary $\theta_s$-eigenoperators  
$e^{-it\hobs_\omega}$ 
in $\agrav$ generating the automorphism $\alpha_t$ 
allows us to apply Landstad's theorem
\cite[Theorem 2]{Landstad1979}\cite[Proposition X.2.6]{TakesakiII}
in order to conclude that $\agrav$ has the structure of a crossed 
product,
\beq
\agrav = \agravss\rtimes_\alpha \mathbb{R}.
\eeq
Furthermore, the theorem implies that $\theta_s$ is the 
generator of the dual automorphism to $\alpha_t$ on 
the crossed product algebra $\agravss\rtimes_\alpha \mathbb{R}$. 
See \cite{AliAhmad:2024wja} for a similar approach using 
Landstad's theorem to identify a crossed-product structure 
on an algebra.  

Although this proves that $\agrav$ is a crossed product algebra, 
it does not immediately determine the properties of $\agravss$, which could
in principle be the trivial algebra $\mathbb{C}\mathbbm{1}$.  
To see that this is not the case, we invoke the structure theorem
for type $\tthr$ algebras as well as Takesaki duality.  
Because $\agrav$ is a type $\ttwo_\infty$ factor,
the structure theorem \cite[Theorem XII.1.1]{TakesakiII} implies
that taking the 
crossed product by the trace-scaling automorphism $\theta_s$ results 
in a type $\tthr_1$ factor $\wh{\alg} = 
\agrav\rtimes_\theta \mathbb{R}$.  However, 
$\agrav$ is itself already a crossed product algebra, and so 
\beq
\wh{\alg} = (\agravss\rtimes_\alpha\mathbb{R})\rtimes_\theta\mathbb{R},
\eeq
with $\theta$ the dual automorphism to $\alpha$.  In such a 
situation, Takesaki duality implies that $\wh{\alg}$ is isomorphic
to the tensor product algebra 
\beq
\wh{\alg} \simeq \agravss\otimes\mc{F}_\infty,
\eeq
where $\mc{F}_\infty = \mc{B}(L^2(\mathbb{R}))$ is the type 
$\tone_\infty$ algebra of all bounded operators on the Hilbert space
$L^2(\mathbb{R})$ \cite{Takesaki1973}\cite[Theorem X.2.3]{TakesakiII}.  Since $\wh{\alg}$ is a type $\tthr_1$ factor
and $\mc{F}_\infty$ is a type $\tone_\infty$ factor, this 
equation implies that $\agravss$ must be a type $\tthr_1$ factor,
which is in fact
isomorphic to $\wh{\alg}$, since then $\agravss\otimes 
\mc{F}_\infty \simeq \agravss$.

The above discussion shows that any flow of trace-scaling automorphisms
$\theta_s$ on a $\ttwo_\infty$ factor must have a fixed point 
algebra that is a type $\tthr_1$ factor.  This contrasts
starkly with the situation for modular automorphisms on 
type $\tthr_1$ factors, where 
depending on the choice of weight $\varphi$,
the centralizer can vary: for integrable weights, centralizers of types $\ttwo_\infty$ and 
type $\ttwo_1$ appeared, and there also 
exist ergodic states with trivial centralizer, such 
as the Minkowski vacuum for a Rindler wedge.  
This difference between modular automorphisms and trace-scaling
automorphisms has a cohomological explanation.  
By Connes's cocycle derivative 
theorem \cite{Connes1973}\cite[Theorem VIII.3.3]{TakesakiII}, 
any two modular flows $\sigma_t^\varphi$, $\sigma_t^\psi$
with respect to the weights $\varphi$ and $\psi$ 
on a type $\tthr_1$ factor 
$\mc{M}$ are outer equivalent, meaning they are related 
by a cocycle perturbation,
\beq \label{eqn:cocycle}
\sigma_t^\varphi = \Ad(\msf{u}_t)\circ \sigma_t^\psi,
\eeq
where $\Ad(\msf{u}_t)$ is the inner automorphism $\msf{u}_t
(\cdot) \msf{u}_t^*$ generated 
by the unitary $\msf{u_t}\in\mc{M}$, and 
the {\it Connes cocycle} $\msf{u}_t$ is a family of 
unitary operators in $\mc{M}$ satisfying the cocycle condition
\beq
\msf{u}_{t+u} = \msf{u}_t \sigma_t^\psi(\msf{u}_u).
\eeq
A  cocycle is called a {\it coboundary} if it can be 
written as
\cite[Section X.1]{TakesakiII}
\beq
\msf{u}_t = \msf{w}\sigma_t^\psi(\msf{w}^*)
\eeq
with $\msf{w}$ a unitary operator in $\mc{M}$.  When
the cocycle relating the flows $\sigma_t^\varphi$ and 
$\sigma_t^\psi$ is a coboundary, the two flows are actually
conjugate, as opposed to only outer conjugate, since 
\begin{align}
\sigma_t^\varphi = \Ad(\msf{w}\sigma_t^\psi(\msf{w}^*))\circ
\sigma_t^\psi
=\Ad(\msf{w})\circ\sigma_t^\psi\circ\Ad(\msf{w}^*),
\end{align}
making use of the identity $\Ad(\sigma_t^\psi(\msf{w}^*))
=\sigma_t^\psi\circ\Ad(\msf{w}^*)\circ\sigma_{-t}^\psi$.
In this case, the centralizers of the two weights are related
by conjugation,
\beq
\mc{M}_\varphi = \msf{w}\mc{M}_\psi\msf{w}^*.
\eeq
Thus, when two weights $\varphi$, $\psi$ have non-conjugate
centralizers, the cocycle $\msf{u}_s$ must not be a coboundary,
meaning that it defines a nontrivial element of the 
cohomology class $H^1_{\sigma^\psi}(\mathbb{R},\mc{U}(\mc{M}))$.  

On the other hand, for a trace-scaling flow $\theta_s$, one 
can prove that every cocycle is a coboundary 
\cite[Theorem XII.1.11]{TakesakiII}, meaning that any flow
that is outer
conjugate $\theta_s$ will have a unitarily equivalent
fixed point algebra.  This is consistent with the 
result discussed above that on a $\ttwo_\infty$ factor,
any trace-scaling flow has a type $\tthr_1$ fixed-point algebra;
there is no way to find an outer-equivalent flow 
with a trivial fixed-point algebra.  Note that in principle,
a given $\ttwo_\infty$ factor could admit distinct 
trace-scaling flows $\theta_s$ and $\eta_s$ that 
are not conjugate, in which case their respective
type $\tthr_1$ 
fixed point algebras would not be isomorphic.  
However, for the hyperfinite $\ttwo_\infty$ algebra, 
the trace-scaling flow is unique up to conjugation,
which follows from the uniqueness of the hyperfinite 
$\tthr_1$ algebra that appears as the fixed point
algebra for the flow \cite{Haagerup1987, Connes1985}.

In order to identify $\agrav$ as a crossed product
algebra, we had to specify a weight $\omega$ on the 
fixed point algebra $\agravss$, which then led to the 
$\theta_s$-eigenoperators $e^{it\hobs_\omega}$. 
The question then arises as to which aspects of the algebra 
$\agrav$ depend on this choice of weight.  Clearly
both $\agrav$ and $\agravss$ are  defined independent
of $\omega$, so the only aspect of $\agrav$ that 
depends on this choice is the explicit parameterization
of $\theta_s$-eigenoperators that, together with $\agravss$,
generate $\agrav$.  Choosing a different weight $\varphi$
on $\agravss$ leads to a different set of eigenoperators
$e^{it\hobs_\varphi}$.  Since $\hobs_\varphi$ must 
generate modular flow with respect to the weight $\varphi$ 
on $\agravss$, we find that the eigenoperators must be 
related by 
\beq
e^{it\hobs_\varphi} = \msf{u}_t e^{it\hobs_\omega},
\eeq
where $\msf{u}_t\in\agravss$ is the Connes cocycle relating the 
modular flows $\sigma_t^\varphi$ and $\sigma_t^\omega$
on $\agravss$ according to (\ref{eqn:cocycle}).  
The fact that one can generate the same 
algebra $\agrav$ from $\agravss$ by adding either sets 
of operators $\{e^{it\hobs_\varphi}\}$ or 
$\{e^{it \hobs_\omega}\}$ points to a kind of 
frame independence of the gravitational algebra 
$\agrav$.  
A similar connection between cocycle perturbations and 
equivalences of the gravitational algebras was noted 
by Witten in his original paper on gravitational
crossed product \cite{Witten2021}.  There, this equivalence
was identified as background independence of the gravitational
algebra, but the above discussion suggests that we 
should instead interpret cocycle equivalences as an independence
of the full algebra on the choice of frame.  
This distinguishes it from the much broader 
notion of background independence later considered 
by Witten in \cite{Witten:2023xze}.  

The frame-dependence inherent to the choice of observer 
has obvious connections to recent works on quantum
reference frames.  Each choice for the observer Hamiltonian
$\hobs$ is associated with a different notion of 
time translation for the shift-symmetric algebra 
$\agrav_0$.  The notion of time provided by the observer
Hamiltonian is actually modular time, with different choices
of states corresponding to different modular time flows
on the algebra.  The connection made here between
the crossed product description and quantum reference
frames is closely related to similar ideas that have 
appeared before \cite{Fewster:2024pur, DeVuyst:2024pop,
AliAhmad:2024wja, DeVuyst:2024uvd}.  We provide
further comments on connections to these works
in section \ref{sec:qrf}, but 
note that one difference is that in the present context,
the observer Hamiltonian is constructed intrinsically
from the quantum fields, and hence the reference frame
is completely intrinsic.  This differs from approaches in
which the reference frame is an external system with 
additional degrees of freedom, and is one of the most interesting
features of the inflationary model.  

\section{Construction of the centralizer} 
\label{sec:constr}
Although the 
identification of $\agrav$ as the image of the time-averaging
operator-valued weight $\Tovw$ 
(\ref{eqn:Tovw}) determines a number of its properties, it is 
somewhat difficult in practice to use this procedure 
to construct explicit operators in $\agrav$.  This is because
the standard local field operators $\phi(x)$, or even 
spacetime smeared versions of them $\int f\phi$, do 
not have finite expectation values in the Bunch-Davies
weight $|\Psi_\text{BD}\rangle$ due to the IR divergence
discussed in section \ref{sec:inflalg}.  The operators on which
$\Tovw$ converges involve nontrivial functions 
of the smeared fields, and the explicit evaluation 
of the time-average integral on these operators is generally
complicated.  Here, we will given an alternative procedure 
for constructing elements of the centralizer, based 
on a perturbation series in the slope of the scalar field
potential $c\ell^2$.  This series gives good approximations 
to the elements of the centralizer when the slope is large.  
At large values of $c\ell^2$, the scalar field rolls down
more quickly, and thus provides a more accurate clock for 
constructing dressed observables.  The corrections to the local
smeared operators generated by the perturbation series
are generically nonlocal, but are small in the limit of large
potential slope.  Hence, we find that the boost-invariant
operators look more local as the clock becomes more 
accurate.

For this construction, we would like to directly 
take advantage of the commutation relation (\ref{eqn:ccr}) by
working with fields smeared only on the $T=0$ Cauchy surface.  
Normally, smearing on a spatial surface is not enough to 
produce well-defined operators with finite fluctuations;
instead, one normally requires operators to be 
smeared in timelike directions \cite{Borchers1964}.  
In the present context, however,
the free scalar field $\phi(x)$ has a low enough scaling dimension
that spatial smearing is sufficient.  Here we will derive the 
precise condition on the smearing function for $\phi(\chi)$ 
and $\pi(\chi)$, and then use these operators to generate the 
algebras of interest.  

We can obtain normalizable states for the scalar field
by acting on the Bunch-Davies weight (\ref{eqn:psiBD}) 
with an operator 
constructed from the scalar field zero mode $f(\varphi_0)$
satisfying the condition that 
\beq
\int_{-\infty}^\infty dy e^{-2\sqrt{2\pi}c\ell^2 y} |f(y)|^2 <\infty.
\eeq
In any such state, the Wightman two-point function of the scalar
field will have a good short-distance behavior, characterized
by the fact that its singular structure has Hadamard form.
In $\text{dS}_2$ near the $T=0$ slice, this condition
reads
\beq \label{eqn:phiphi}
\langle \phi(T,\chi_1) \phi(S, \chi_2)\rangle = 
\log\left[-(T-S-i\vep)^2 + (\chi_1-\chi_2)^2\right] + \text{finite}
\eeq
where the limit $\vep\rightarrow 0$ defines this 
two-point function as a distribution for real values 
of $T,S$.  Since we are only interested in the possible
singularities in this two-point function at coincident
points, we will drop the finite terms and simply analyze the 
behavior of smeared operators coming only from the $\log$ term. 
On the $T=S=0$ slice, this becomes
\beq
\langle\phi(\chi_1) \phi(\chi_2)\rangle = 
\log\left[(\chi_1-\chi_2)^2 + \vep^2\right]
\eeq
We also want to analyze the $\pi(\chi)$ two-point function, 
we we can obtain from (\ref{eqn:phiphi}) by taking $T$ and $S$
derivatives (note that no contact terms appear
because this is a Wightman function, as opposed to 
a time-ordered correlation function),
\begin{align}
\langle \pi(T,\chi_1) \pi(S,\chi_2)\rangle
&=\partial_T \partial_S \log\left[-(T-S-i\vep)^2 + (\chi_1-\chi_2)^2\right] \nonumber \\
&= 
\partial_{\chi_1} \partial_{\chi_2} \log\left[-(T-S-i\vep)^2 + (\chi_1-\chi_2)^2\right] \nonumber \\
&\overset{T,S=0}{=}
\partial_{\chi_1} \partial_{\chi_2} \log\left[ (\chi_1-\chi_2)^2 + \vep^2\right].  \label{eqn:pipi}
\end{align}

Using these, we can determine the condition on the 
spatially smeared field operators 
$\phi_f = \int d\chi f(\chi) \phi(0,\chi)$, 
$\pi_g = \int d\chi g(\chi) \pi(0,\chi)$ to ensure 
that the operators have finite fluctuations.  
Beginning with $\phi_f$, the smeared two-point function
is (with $\chi_{12} = \chi_1-\chi_2$)
\begin{align}
\langle \phi_f^2\rangle 
&= 
\int d\chi_1 \int d\chi_2 f(\chi_1)
f(\chi_2) \log(\vep^2+\chi_{12}^2) \nonumber \\
&=
\int d\chi_2 f(\chi_2) \int d\chi_{12} f(\chi_{12}+\chi_2)
\log(\vep^2 + \chi_{12}^2)
\end{align}
The possible divergence clearly comes from $\chi_{12}$ near zero,
and by Taylor expanding $f(\chi_2 +\chi_{12})$ near $\chi_2$,
only the leading term in the expansion $f(\chi_2)$ appears
multiplying a divergent integrand as $\vep\rightarrow 0$.  
To check that this contribution is finite as $\vep\rightarrow 0$,
we can simply integrate $\log(\chi_{12} +\vep^2)$ between
$(-\lambda,\lambda)$, where $\lambda$ is small but finite.  
This gives
\beq
\int_{-\lambda}^\lambda d\chi_{12} \log(\chi_{12}^2 +\vep^2)
= 4\lambda(\log(\lambda)-1) + \op(\vep),
\eeq
which is therefore finite as $\vep\rightarrow 0$.  Hence, as 
long as the smearing function $f(\chi)$ is finite everywhere,
we obtain an operator with finite fluctuations.  In particular,
$f(\chi)$ need not be continuous, much less smooth.  

The smeared $\pi(\chi)$ operator has a higher dimension than
$\phi(\chi)$, so we should expect a stronger divergence in the 
two-point function.  However, the particular form 
of the two-point function (\ref{eqn:pipi}) involving 
spatial derivatives will result in a class of spatial smearings 
that work.  We take the smearing 
$g(\chi)$ to be supported 
between $\chi = a$ and $\chi = b$, and find
\begin{align}
\langle \pi_g^2\rangle
&=\int_a^b d\chi_1 \int_a^b d\chi_2 
g(\chi_1) g(\chi_2)\partial_{\chi_1}
\partial_{\chi_2} \log (\chi_{12}^2 + \vep^2)
\nonumber \\
&= g(\chi_1) g(\chi_2)\log(\chi_{12}^2 +\vep^2)\Big|_{\chi_1=a}^b
\Big|_{\chi_2=a}^b \nonumber \\
&\quad -2\int_a^b d\chi_1 g(\chi_2) g'(\chi_1) \log(\chi_{12}^2 + \vep^2)
\Big|_{\chi_2 = a}^b
+\int_{a}^b d\chi_1 \int_{a}^b d\chi_2 g'(\chi_1) g'(\chi_2) 
\log(\chi_{12}^2 +\vep^2).
\end{align}
By a similar argument as before 
the integrals in the second line are finite as long as $g(\chi)$ and 
$g'(\chi)$ are finite, hence we require $g(\chi)$ to be differentiable almost everywhere
on the interior of its support.  The first line contains the possible
divergences when $\chi_1 = \chi_2 = a$ or $\chi_1 = \chi_2 = b$, proportional
to $g(a)^2 \log\vep^2$, $g(b)^2\log(\vep^2)$.  To avoid these, 
we require that $g(\chi)$ vanish on the boundary 
of its support.  This condition in fact follows from the 
differentiability of $g(\chi)$, since this implies that $g(\chi)$
is continuous.  Since it vanishes by definition on the complement
of the support, by continuity it also vanishes on the 
boundary.\footnote{One could also try to weaken
the restrictions on $g(\chi)$ by adding 
boundary terms to the smeared $\pi_g$ operator 
at points where $g(\chi)$ is not 
differentiable.  An operator with the 
correct dimension is just $\phi(\chi)$, and one might attempt 
to cancel the divergent fluctuations in $\pi_g$ from nondifferentiable
points of $g$ against the local fluctuations 
in $\phi(\chi)$, e.g. with an operator of the form
\beq \label{eqn:piimp}
\pi_g + \alpha \phi(a)
\eeq
in the case where 
$\chi=a$ is the only point where $g(\chi)$ is discontinuous.  
This ends up not working because the divergences coming from 
$\pi_g^2$ and $\phi(a)^2$ both have positive coefficients
respectively proportional to $g(a)^2$ and 
$\alpha^2$, and hence cannot cancel each other, while the 
contribution coming from the cross terms $\langle \pi_g \phi(a)\rangle$
and $\langle\phi(a) \pi_g\rangle$ are finite (if one uses
complex $\alpha$ and $g$, the same issue would 
arise when computing $\langle (\pi_g + \alpha\phi(a))^*(\pi_g+\alpha\phi(a))\rangle$, with 
divergences involving $|g(a)|^2$ and $|\alpha|^2$).  
This is because the 
short distance behavior of $\langle \pi(\chi) \phi(a)\rangle$, obtained 
by taking a time derivative of (\ref{eqn:phiphi}), 
goes like
\beq
\langle \pi(\chi) \phi(a)\rangle \sim \frac{2i\vep}{(\chi-a)^2 + \vep^2},
\eeq
which yields only finite contributions  as $\vep\rightarrow 0$ when
integrating $\chi$ around $\chi = a$.  Since these terms 
then cannot cancel the divergence, we see that any improved operator 
of the form (\ref{eqn:piimp}) also has divergent fluctuations. 
}
In higher dimensions, similar arguments can be used to show that 
spatial smearing with a bounded 
function $f(x)$ is sufficient 
for the field $\phi(x)$, and a differentiable function 
for the momentum  $\pi(x)$.  Hence the procedure described in 
this section has straightforward generalizations
to higher dimensional constructions.  

The local static patch algebra is then generated by products 
of the spatially smeared operators $\phi_f$ and $\pi_g$, 
where both $f$ and $g$ are supported in the static patch, $f$ is 
finite everywhere but not necessarily continuous, and $g$ is continuous
and
differentiable everywhere (although $g'$ need not be continuous).
To find the boost-invariant operators, we write the boost 
generator (\ref{eqn:Hxi}) as 
\begin{align}
H_\xi &= H_0 + c\ell^2(\hat{x} - \hat{x}') \\
H_0 &= \int d\chi \cos\chi\left(\frac12 \pi(\chi)^2 + \frac12
\phi'(\chi)^2\right) \\
\hat{x} &= \int_{-\frac{\pi}{2}}^{\frac{\pi}{2}} d\chi \cos\chi \phi(\chi) \\
\hat{x}' &= -\int_{\frac{\pi}{2}}^{\frac{3\pi}{2}} \cos\chi \phi(\chi).
\end{align}
By the above discussion, $\hat{x}$ is a well-defined 
operator affiliated with the static patch algebra, while $\hat{x}'$
is affiliated with the commutant algebra $\alg'$.  $H_0$ is the boost 
generator for the standard massless scalar field with no potential.  
This operator also generates a modular flow on $\alg$ associated with the 
Bunch-Davies weight of the scalar with zero potential.  The modular 
flow associated to $H_0$ is expected to act ergodically on 
$\alg$, so we should not expect to find any operators that commute 
with $H_0$.  

The operators $\hat{x}$ and $\hat{x}'$ are then related 
to the Connes cocycles between the two Bunch-Davies weights 
$\omega_\text{BD}^c$ and $\omega_\text{BD}^0$.   The cocycle
associated with the algebra $\alg$ can be defined in terms of the 
relative modular operators according to
\beq \label{eqn:coc}
\msf{u}_{c|0}(s) = \Delta_c^{is} \Delta_{0|c}^{-is},
\eeq
and it determines the relation between the modular flows of the two
weights via\footnote{The minus sign in $\msf{u}_{c|0}(-t)$ is appearing
because we have defined the forward time direction for modular flow
$\sigma_t$ opposite to the standard choice in mathematics literature, i.e.\
our definition is 
$\sigma_t(\msf{a}) = \Delta^{-it} \msf{a} \Delta^{it} = e^{ith}\msf{a}
e^{-ith}$.  This choice is more convenient since modular flow 
then satisfies a KMS condition with positive temperature, whereas
the math convention leads to a KMS condition with negative temperature.  
We have kept the definition of the cocycle (\ref{eqn:coc}) the same as
in mathematics literature, which then results in the minus sign in 
(\ref{eqn:sigcsig0}).}
\beq \label{eqn:sigcsig0}
\sigma_t^c = \Ad(\msf{u}_{c|0}(-t))\circ \sigma_t^0.
\eeq
The operator $\hat{x}$ is then just the leading order piece of the 
cocycle at small $s$:
\beq
\msf{u}_{c|0}(-s) = \mathbbm{1} + is2\pi \hat{x} + \ldots.
\eeq
Similarly, the operator $\hat{x}'$ is related to the cocycle
for the commutant algebra,
\beq
\msf{u}_{0|c}'(s) = \Delta_0^{-is}\Delta_{0|\xi}^{is}, 
\qquad \msf{u}_{0|c}'(-s) = \mathbbm{1} - is2\pi\hat{x}' + \ldots.
\eeq

Since all operators in $\alg$ commute with $\hat{x}'$, the problem
reduces to finding operators in $\alg$ that commute
with $H_0 + c\ell^2 \hat{x}$.  This can be solved recursively
as a power series in $\frac{1}{c\ell^2}$.  We express the operator
$\tilde{\msf{a}}$ affiliated with the centralizer as 
\beq\label{eqn:atilsum}
\tilde{\msf{a}} = \sum_{n=0}^\infty \frac{1}{(c\ell^2)^n} \msf{a}_n.
\eeq
The condition that $\tilde{\msf{a}}$ commutes with $H_0+c\ell^2\hat{x}$
 then translates to the following
recursion relation on operators $\msf{a}_n$:
\beq \label{eqn:rec}
[\hat{x}, \msf{a}_n] = \begin{cases} 0 & n=0 \\ -[H_0,\msf{a}_{n-1}] & n\geq 1
\end{cases}
\eeq

This  leads to an algorithm for perturbatively solving for the 
operators $\msf{a}_n$.  It is straightforward to parameterize
the operators solving the initial condition $[\hat{x},\msf{a}_0] = 0$:
$\hat{x}$ is linear in the field operator $\phi(\chi)$, and it commutes 
with all spatially smeared field operators $\phi_f$ affiliated with
$\alg$, as well
as spatially smeared momentum operators $\pi_g$ subject to the condition
\beq
\int_{-\frac{\pi}{2}}^{\frac{\pi}{2}} d\chi \cos(\chi) g(\chi) = 0
\eeq
A generic operator affiliated with
$\alg$ commuting with $\hat{x}$ will then be 
expressible as a sum of products of these smeared single-field 
operators $\phi_f$ and $\pi_g$.  

Since $H_0$ is quadratic in the field operators, its 
 action on such a product of smeared field operators
is straightforward to determine using the commutation
relation (\ref{eqn:ccr}) and the Leibniz rule.  For example,
\begin{align}
[H_0,\pi_g] &= \int d\chi_1 d\chi_2 g(\chi_2)\cos(\chi_1)\phi'(\chi_1)[\phi'(\chi_1),\pi(\chi_2)] \nonumber \\
& =
i\int d\chi_1 d\chi_2 g(\chi_2) \cos(\chi_1) \phi'(\chi_1)
\partial_{\chi_1}\delta(\chi_1-\chi_2) \nonumber \\
&=
-i\int d\chi g(\chi) \partial_\chi(\cos(\chi) \phi'(\chi)) 
\nonumber \\
&= -i \phi_{(g'\cos(\chi))'}
\end{align}
On an operator involving $j$ single-field operators of the form $\psi_{f_1}\psi_{f_2}\cdots \psi_{f_j}$ 
where $\psi$ is either $\phi$ or $\pi$ and $f_i$ are appropriate
smearings, the commutator with $H_0$ will result in a sum 
of terms involving at most $j$ single field operators.  
Hence the right hand side of the recursion (\ref{eqn:rec}) can
always be computed.  

Finally, we need to check that given a multi-field operator 
$\msf{b}_j =\psi_{f_1}\cdots \psi_{f_j}$, we can always find a new operator  ${\msf{c}}_{j+1}$ with $\hat{x}$ satisfying $[\hat{x}, \msf{c}_{j+1}] = 
\msf{b}_j$.  Using the canonical commutation relation (\ref{eqn:ccr})
we can always express $\msf{b}_j$ as a finite sum of terms of the form
\beq \label{eqn:pign}
(\pi_{\hat{g}})^n \tilde\psi_{f_1} \cdots \tilde\psi_{f_{j-m-n}}
\eeq
where the $\tilde{\psi}_{f_k}$ are all single-field operators 
that commute with $\hat{x}$,  and $m\geq 0$, so that all terms involve
at most $j$ single-field smeared operators.  The function $\hat{g}$ 
is chosen to be $\hat{g}(\chi) = \frac{4}{\pi} \cos(\chi)\Theta(|\chi|-\frac{\pi}{2})$, with $\Theta$ the Heaviside step function, so 
that $\pi_{\hat{g}}$ is a canonical conjugate to $\hat{x}$
satisfying $[\hat{x}, \pi_{\hat{g}}] = i$.   Then the 
operator appearing in (\ref{eqn:pign}) can be written as a commutator
with $\hat{x}$,
\beq
\left[\hat{x}, \frac{-i}{n+1}(\pi_{\hat{g}})^{n+1} \tilde\psi_{f_1}
\cdots \tilde\psi_{f_{j-m-n}}\right] 
= (\pi_{\hat{g}})^n \tilde\psi_{f_1} \cdots \tilde\psi_{f_{j-m-n}}.
\eeq
This demonstrates that, given the 
operator $\msf{b}_{n} = -[H_0, \msf{a}_{n-1}]$ appearing in the 
recursion 
relation (\ref{eqn:rec}), we can always find $\msf{a}_n$ satisfying
$[\hat{x}, \msf{a}_n] = \msf{b}_n$.  There are in fact many 
solutions to this relation, since given any one solution, 
we can always add an operator commuting with $\hat{x}$. The plethora of 
solutions found at each step in the recursion 
relation leads to many operators $\tilde{\msf{a}}$, constructed 
as in
(\ref{eqn:atilsum}),  that formally commutes
with $H_0 + c\ell^2 \hat{x}$.  To make this procedure precise,
one would also have to check that the sum (\ref{eqn:atilsum})
produces an operator with finite fluctuations, and this may impose
some constraints on how to choose the $\msf{a}_n$ at each step
in the recursion.  We do not examine these constraints in detail here, but 
it seems clear that there will be many choices that result in well-defined
operators $\tilde{\msf{a}}$ affiliated with the centralizer.

Note that each term in the sum (\ref{eqn:atilsum}) is suppressed 
by a power of $\frac{1}{c\ell^2}$.  At large value of the scalar
field slope $c\ell^2\gg1$, the corrections to the leading terms 
in the sum become increasingly suppressed.  It is also the case 
that each term in the sum is more nonlocal than the preceding term, in the 
sense that if $\msf{a}_n$ involves the product of $k$ integrated field
operators $\psi_{f_i}$, any solution for $\msf{a}_{n+1}$ must involve
products of at 
least $k+1$ $\psi_{f_i}$'s, since the $\hat{x}$ commutator
decreases the number $\psi_{f_i}$ appearing in a product,
while $H_0$ preserves the number.  This leads to the conclusion
that the nonlocality in the operators induced by the requirement
that they commute with the boost generator is suppressed
when the slope $c\ell^2$ is large.  This limit is associated with 
a more quickly rolling scalar field, in which case it provides a more 
accurate clock for dressing the operator in the static patch.  This 
increased accuracy appears quantitatively in the fact that 
the nonlocal corrections to an operator $\msf{a}_0$ that make it boost 
invariant are suppressed by powers of $\frac{1}{c\ell^2}$. 

In the limit of small slope $c\ell^2\rightarrow 0$, this perturbative 
procedure breaks down, because each term in the sum (\ref{eqn:atilsum})
is enhanced relative to the previous one, suggesting that it will
not converge to an operator with finite fluctuations.  This is in line
with the expectation that the zero-slope modular Hamiltonian $2\pi H_0$ 
acts ergodically, so that no operators localized to the static
patch commute with it.  

Via a slight specialization of the above procedure, one can construct
the observer Hamiltonian $\hobs$ guaranteed to exist in the crossed 
product description of $\agrav$ according to the discussion
of section \ref{sec:crossprod}. This operator is defined by the 
relations 
\beq \label{eqn:hobsreln}
[H_\xi , \hobs] = 0, \qquad [\Pi, \hobs] = i
\eeq
where $\Pi$ is the generator of the shift symmetry
defined in (\ref{eqn:Pi}).  As above, we write the 
operator $\hobs$ as a power series in $\frac{1}{c\ell^2}$,
\beq
\hobs = \sum_{n=0}^\infty \frac{1}{(c\ell^2)^n}\vep_n.
\eeq
For the initial operator $\vep_0$, we choose a 
canonical conjugate to $\Pi$, the simplest of 
which is the constant smeared field operator,
\beq
\vep_0 = 4c\ell^2\int_{-\frac{\pi}{2}}^{\frac{\pi}{2}}
d\chi \phi(\chi).
\eeq
We then solve for the subsequent operators $\vep_n$, subject 
to the additional requirement that $[\Pi,\vep_n] = 0$
for $n\geq 1$.  This requirement implements the 
second condition in (\ref{eqn:hobsreln}), since 
then $[\Pi, \hobs] = [\Pi, \vep_0] = i$.   

It remains 
to check that this additional condition can always
be satisfied.  First note that because
\beq
[H_0, \vep_0] = -i 4c\ell^2 \int_{-\frac{\pi}{2}}^{\frac{\pi}{2}}
d\chi \cos(\chi) \pi(\chi),
\eeq
it holds that $[\Pi,[H_0,\vep_0]] = 0$.  Additionally, 
for $n\geq 1$, if we have shown that $\vep_n$ commutes
with $\Pi$, then we also have that $[\Pi,[H_0,\vep_n]] = 0$,
since $[\Pi, H_0] = 0$.  We therefore have to 
show that whenever $[\Pi,[H_0,\vep_n]] = 0$,
we can find a solution $\vep_{n+1}$ to the recursion
relation (\ref{eqn:rec}) satisfying $[\Pi,\vep_{n+1}] = 0$.  
Take $\msf{b}_{n+1}$ to be a generic solution
to the relation, so that 
$[\hat{x},\msf{b}_{n+1}] = -[H_0,\vep_n]$.  Then
since $[\Pi,\hat{x}] = 0$, we find that 
\beq
0 =[\Pi,[\hat{x},\msf{b}_{n+1}]] = [\hat{x},[\Pi,\msf{b}_{n+1}]].
\eeq
This shows that although an arbitrary solution $\msf{b}_{n+1}$
may not commute with $\Pi$, the commutator with 
$\Pi$ commutes with $\hat{x}$.  A generic operator 
commuting with $\hat{x}$ can be written as a 
sum of terms of the form $\vep_0^k \msf{a}_k$,
with $\vep_0^k$ the part that does not commute
with $\Pi$, and $\msf{a}_k$ an operator that commutes 
with $\Pi$.  We therefore have the relation
\beq
[\Pi, \msf{b}_{n+1}] = \sum_k \vep_0^k \msf{a}_k,
\eeq
which is solved by
\beq
\msf{b}_{n+1} = \sum_k  \left(\frac{-i}{k+1}
\vep_0^{k+1}\msf{a}_k +
\msf{c}_k\right)
\eeq
where $[\Pi,\msf{c}_k]=0$.  The terms $\vep_0^{k+1}\msf{a}_k$
all commute with $\hat{x}$, and hence they 
can be subtracted from $\msf{b}_{n+1}$ to obtain
another solution to the recursion 
relation.  Hence, the operator
\beq
\vep_{n+1} = \msf{b}_{n+1} - \sum_k\frac{-i}{k+1}\vep_0^{k+1}
\msf{a}_k = \sum_k \msf{c}_k,
\eeq
gives the desired solution to the recursion 
relation that also commutes with $\Pi$.  

The remaining ambiguity in the choice of $\vep_n$ at
each step is an operator that commutes with both $\Pi$
and $\hat{x}$.  There are many such operators; 
the full algebra of them is 
generated by field operators 
$\phi_f$ with $\int d\chi f(\chi)=0$ and momentum
operators $\pi_g$ with $\int d\chi \cos(\chi) g(\chi)  = 0$.  
As discussed in section \ref{sec:crossprod}, each such solution
will correspond to a different weight on the shift-symmetric
subalgebra $\agrav_0$, for which $\hat{\vep}$ is the 
generator of modular flow.

\section{Time operator}
\label{sec:time}

Although we showed in section \ref{sec:crossprod} that $\agrav$ 
has a canonical description as a crossed product, the presentation
of the algebra looks different from crossed products that 
have appeared recently in other works on gravitational
algebras \cite{Witten2021, Chandrasekaran2022a, Chandrasekaran2022b,
Jensen2023, Kudler-Flam:2023qfl, Faulkner:2024gst}.  
In these descriptions,
the kinematical algebra possesses a natural tensor factorization 
into a factor associated with quantum fields and a factor associated 
with the observer which contains the observer Hamiltonian and a canonically
conjugate time operator.  The discussion of section \ref{sec:crossprod}
showed that the observer Hamiltonian $\hobs$ must exist, and in 
section \ref{sec:constr} we gave a perturbative construction
of this operator.  Hence the main obstruction to matching 
the present inflationary example to previous crossed
product constructions is the identification of a time
operator in the kinematical algebra $\alg$.  In this section,
we describe the properties that such an operator should have, 
and attempt to determine this operator in $\alg$.  However, 
despite various uniqueness results involve trace-scaling 
automorphisms of hyperfinite factors, we will
find that there does not appear to be an operator in the kinematical
algebra that generates the action of the automorphism $\theta_s$ on 
$\agrav$. We will show that there is a different 
class of trace-scaling automorphisms of $\agrav$ that do
arise from time operators in $\alg$, but none of them appear to be 
canonically preferred, unlike the shift symmetry generator $\Pi$.  
This lack of preferred time operator is therefore one of the 
central differences between the inflationary gravitational 
algebra and other crossed product constructions.

For this discussion, we will define a time operator $\tp$ as 
an operator affiliated with the kinematical algebra $\alg$
that is canonically conjugate to the Bunch-Davies modular 
Hamiltonian $h = 2\pi H_\xi$,
\beq \label{eqn:ht}
[h, \tp] = -i.
\eeq
Such an operator is guaranteed to exist 
when $H_\xi$ is the generator of modular flow of a dominant 
weight on $\alg$.  To see why, 
recall that $\agrav$ is the centralizer 
of this dominant weight and $\alg$ is isomorphic
to a crossed product algebra,
\beq
\alg \simeq \agrav\rtimes_\eta\mathbb{R}
\eeq
where $\eta_s$ is a particular trace-scaling flow of automorphisms
of $\agrav$.  
There then must be a one-parameter group 
of unitary operators 
$\lambda(s) \in \alg$ implementing $\eta_s$ on $\agrav$
by conjugation:
\beq
\lambda(s) \tilde{\msf{a}} \lambda(s)^* = \eta_s(\tilde{\msf{a}}),
\quad \tilde{\msf{a}}\in\agrav.
\eeq
These $\lambda(s)$ define a family of inner automorphisms 
of the kinematical algebra $\alg$ that commutes with the 
modular flow $\sigma_t$ of the Bunch-Davies weight.  
Because of this, the automorphism $\Ad(\lambda(s))$ commutes
with the operator-valued weight $\Tovw$ defined by (\ref{eqn:Tovw}).
Furthermore since $\omega_\text{BD} = \tau_\text{BD}\circ \Tovw$
and because $\eta_s$ rescales the trace $\tau_\text{BD}$, 
we find that the automorphism $\Ad(\lambda(s))$ rescales the 
Bunch-Davies weight,
\beq
\omega_\text{BD}\circ \Ad(\lambda(s)) = e^{-s} \omega_\text{BD}.
\eeq
In fact, the existence of such unitary operators in $\alg$ rescaling
the weight $\omega_\text{BD}$ is the defining 
property of a dominant weight \cite[Theorem II.1.1]{Connes1977}\cite[Theorem XII.4.18]{TakesakiII}.  

Since the modular automorphism $\sigma_t$ is the dual
automorphism associated with the 
realization of $\alg$ as an $\eta_s$-crossed product,
the operators $\lambda(s)$ are unitary eigenoperators for the 
modular flow, satisfying 
\beq \label{eqn:sigteigen}
\sigma_t(\lambda(s)) = e^{ith} \lambda(s) e^{-ith} = e^{ist} \lambda(s).
\eeq
Hence, writing $\lambda(s) = e^{is\hat{t}}$, this relation 
shows that $\hat{t}$ is conjugate to $h$, satisfying 
the commutation relation (\ref{eqn:ht}).

Next we discuss the relation between $\eta_s$ and 
the trace-scaling automorphism $\theta_s$ associated 
with the massless field shift symmetry.  An important
point here is that because $\agrav$
is a hyperfinite $\ttwo_\infty$ factor, flows
of trace-scaling
automorphisms are unique up to conjugation (see 
appendix \ref{sec:automorphism}), meaning
that there is some $\alpha \in \aut(\agrav)$ such 
that 
\beq\label{eqn:thconj}
\theta_s = \alpha\circ\eta_s \circ \alpha^{-1}.
\eeq
Crossed products by conjugate flows are isomorphic
\cite[Theorem X.1.7]{TakesakiII},
and the isomorphism $\Phi:\agrav\rtimes_\theta \mathbb{R}
\rightarrow \alg$ acts by 
\beq\label{eqn:etaconj}
\Phi(\tilde{\msf{a}}) = \alpha^{-1}(\tilde{\msf{a}}),\quad
\Phi(\mu(s)) = \lambda(s),
\eeq
where $\mu(s)$ are the unitary operators in $\agrav\rtimes_\theta
\mathbb{R}$ generating the action of $\theta_s$.  

Unfortunately, this is not enough to conclude that there 
are operators in $\alg$ that implement 
the automorphism $\theta_s$ on its $\agrav$ subalgebra. 
If $\alpha$ in equation (\ref{eqn:etaconj}) could be chosen
such that it lifts to an automorphism of $\alg$, this would 
lead to the desired result, since then the operators 
$\alpha(\lambda(s))$ would generate $\theta_s$ 
on the $\agrav$ subalgebra:
\beq
\alpha(\lambda(s)) \tilde{\msf{a}}\alpha(\lambda(s)^*)
=
\alpha(\lambda(s)\alpha^{-1}(\tilde{\msf{a}})\lambda(s)^*)
=
\alpha\circ\eta_s\circ\alpha^{-1}(\tilde{\msf{a}})
=
\theta_s(\tilde{\msf{a}}).
\eeq
Conversely, if $\theta_s$ lifts to a family of inner 
automorphisms on $\alg$, we can again apply Landstad's theorem
\cite[Theorem 2]{Landstad1979}\cite[Proposition X.2.6]{TakesakiII}
to conclude that $\alg$ is canonically an $\theta_s$-crossed
product, meaning that the $\Phi$ in (\ref{eqn:etaconj})
defines an automorphism of $\alg$.  

This raises the question as to which automorphisms $\alpha \in
\aut(\agrav)$ lift to automorphisms of $\alg 
= \agrav \rtimes_\eta \mathbb{R}$.  Since any inner
automorphism of $\agrav$ lifts trivially to an
inner automorphisms of $\alg$, we need only characterize the 
subgroup of the outer automorphism group $\out(\agrav)$ 
coming from automorphisms that lift to $\alg$.  Using, for example,
the arguments in the proof of Theorem XII.1.10(iv) in 
\cite{TakesakiII} as well as the uniqueness of the trace-scaling
automorphism $\eta_s$ up to conjugation, we find that 
the relevant subgroup is $\out_\eta(\agrav)$, consisting 
of automorphisms that commute with $\eta_s$ modulo inner 
automorphisms.  This group has a direct product structure
\beq \label{eqn:outeta}
\out_\eta(\agrav) = \out_{\eta,\tau}(\agrav)\times \mathbb{R},
\eeq
where $\out_{\eta,\tau}(\agrav)$ 
is the subgroup of trace-preserving 
automorphisms in $\out_\eta(\agrav)$, 
and the factor $\mathbb{R}$ is the trace-scaling
automorphism generated by $\eta_s$.  
Without loss of generality we can choose 
$\alpha$ in (\ref{eqn:thconj}) to be trace-preserving
by making an appropriate shift $\alpha \rightarrow \alpha\circ
\eta_{s_0}$ for some $s_0$.  Then if $\alpha$ lifts to an
automorphism of $\alg$, its image in $\out(\agrav)$ must 
lie in $\out_{\eta,\tau}(\agrav)$, and hence
there exists some unitary $\tilde{\msf{v}}\in\agrav$ such
that $\Ad(\tilde{\msf{v}})\circ\alpha$ commutes with $\eta_s$.  
This implies that 
\beq
\theta_s = \Ad(\tilde{\msf{v}}^*)\circ \eta_s\circ 
\Ad(\tilde{\msf{v}}) = \alpha\circ \eta_s\circ \alpha^{-1},
\eeq
so in particular $\theta_s$ is related to $\eta_s$ by 
conjugation by an inner automorphism $\Ad(\tilde{\msf{v}}^*)$. 
We can then define an $\eta_s$-cocycle $\tilde{\msf{c}}_s 
= \tilde{\msf{v}}^*
\eta_s(\tilde{\msf{v}})$, in terms of which $\theta_s$ 
can be written 
\beq
\theta_s = \Ad(\tilde{\msf{c}}_s) \circ \eta_s,
\eeq
showing in this case that it is cocycle-equivalent to $\eta_s$,
and hence has the same image as $\eta_s$ in the outer automorphism 
group $\out(\agrav)$.  

This argument additionally can be used to prove the existence
of trace-scaling flows on $\agrav$
that do not arise from inner
automorphisms of $\alg$.  Choose any flow $\beta_s \in
\aut_{\eta,\tau}(\agrav)$ of trace-preserving automorphisms 
commuting with $\eta_s$.  Then $\theta_s = \beta_s \circ \eta_s$ 
is a trace-scaling flow that lifts to a flow of automorphisms
on $\alg$.  
However, its image in $\out(\agrav)$ differs from that 
of $\eta_s$ unless $\beta_s$ is inner.  As argued above,
whenever $\beta_s$ is not inner, the automorphism $\alpha$ 
relating the flows $\eta_s$ and $\theta_s$ in (\ref{eqn:thconj})
cannot be chosen to lift to an automorphism of $\alg$, and 
hence $\theta_s$ must arise from a flow of outer automorphisms
in $\alg$.  This example gives a complete characterization
of the possible trace-scaling automorphisms of $\agrav$:
any such $\theta_s$ can be written as 
\beq
\theta_s = \Ad(\tilde{\msf{c}}_s)\circ \beta_s \circ\eta_s
\eeq
with $\tilde{\msf{c}}_s$ a unitary in $\agrav$, and $\beta_s
\in \aut_{\eta,\tau}(\agrav)$.  We will 
assume $\tilde{\msf{c}}_s = \mathbbm{1}$ 
since we already argued that perturbing 
by inner automorphisms does not affect the ability to 
lift $\theta_s$ to an inner automorphism of $\alg$.  
Hence, $\beta_s$ represents
the true obstruction to lifting $\theta_s$.

The obstruction $\beta_s$ has an interesting interpretation in 
terms of the fixed point algebra $\agravss = \agrav^\theta$.  
Because $\beta_s$ commutes with $\eta_s$, it also commutes
with $\theta_s$, and hence $\beta_s$ defines an automorphism
of $\agravss$.  In fact, any automorphism of $\agravss$ 
 lifts to a trace-preserving automorphism 
of $\agrav$ commuting with $\theta_s$, and the 
groups $\aut(\agravss)$ and $\aut_{\theta,\tau}(\agrav)$ 
are isomorphic \cite{Haagerup1990b}\cite[Exercise XII.1]{TakesakiII}.  This contrasts with the analogous 
situation on the relation between $\agrav$ and $\alg$, 
where only a subgroup of automorphisms of $\agrav$ 
lifts to automorphisms of $\alg$.  Hence, if $\theta_s$ itself
does not lift to an inner automorphism of $\alg$, it must 
be a combination of an outer automorphism $\beta_s$
of the quantum
field theory subalgebra $\agravss$ and the trace-scaling
flow $\eta_s$ which is realized as an inner automorphism
of $\alg$. An example of a possible choice of $\beta_s$ 
to keep in mind is a rotation transformation
of the de Sitter static patch about some axis in spacetime
dimensions $D\geq 2$.

We now argue that the shift symmetry generator $\Pi$ defines 
an outer automorphism of the kinematical algebra $\alg$,
thereby precluding the existence of a time operator associated 
with this automorphism.  If $\Pi$ generated an inner automorphism
of $\alg$, it would be possible to split the generator 
$\Pi = \Pi_\alg + \Pi_{\alg'}$, with $\Pi_\alg$ affiliated with $\alg$
and $\Pi_{\alg'}$ affiliated with the commutant.  It is clear what 
these two operators should be: $\Pi$ is given by an integral of $\pi(\chi)$ 
over the full $T=0$ Cauchy slice with constant smearing, and hence 
$\Pi_{\alg}, \Pi_{\alg'}$ should be given by the component of these 
integrals restricted either to the static patch or its complement,
i.e.\
\beq
\Pi_\alg = -\frac{1}{4\pi c\ell^2}\int_{-\frac{\pi}{2}}^{\frac{\pi}{2}}
d\chi \pi(\chi).
\eeq
However, we argued in section \ref{sec:constr} that in order for 
a spatial smearing of $\pi(\chi)$ to define a good operator 
with finite fluctuations, the smearing must vanish on the boundary
of its support.  The step-function smearing 
$\Theta(|\chi|-\frac{\pi}{2})$ is discontinuous at its boundary,
and so
$\Pi_\alg$ does not have finite fluctuations.   Since it is not
possible to split $\Pi$ into well-defined operators affiliated with 
$\alg$ and $\alg'$, it defines an outer automorphism 
of $\alg$.  

Although $\theta_s$ is not generated by a time operator 
in $\alg$, the above discussion indicates that there is a 
different trace-scaling automorphism $\eta_s$ generated by 
a time operator.  We can in fact solve for this time 
operator using a similar procedure to the one described 
in section \ref{sec:constr}.  We write $\hat{t}$ as an expansion
in $\frac{1}{c\ell^2}$,
\beq
\hat{t} = \sum_{n=0}^\infty \frac{1}{(c\ell^2)^n} t_n,
\eeq
and impose the relations 
\beq
[\hat{x},t_n] = \begin{cases}
\frac{-i}{2\pi c\ell^2} & n=0 \\
-[H_0, t_{n-1}] & n\geq 1
\end{cases}.
\eeq
These relations then ensure that 
\beq
[2\pi H_\xi, \hat{t} ] = [2\pi c\ell^2 \hat{x}, t_0] = -i.
\eeq

While this procedure always results in a time operator $\hat{t}$, it is 
by no means unique.  There is considerable ambiguity in choosing 
the initial canonical conjugate $t_0$ to $\hat{x}$, and as in 
section \ref{sec:constr}, the solution to the recursion relation
at each step is ambiguous up to operators that commute with 
$\hat{x}$.  These ambiguities will result in different time 
operators related by operators in $\agrav$, and this is precisely
the ambiguity present in $\eta_s$ discussed above.  We conclude
that there does not appear to be a preferred time operator in the 
kinematical algebra $\alg$.  

This raises the question of how to interpret this lack of a canonical
time operator in the kinematical algebra $\alg$.  One possible interpretation
is that the lack of a canonical time operator indicates further 
ambiguity in decomposing the system in to an observer algebra 
and a quantum field algebra.  Hence, the choice of a preferred 
inner automorphism $\eta_s$ of $\alg$ indicates additional frame
dependence.  On the other hand, the kinematical algebra $\alg$ is not
necessarily an algebra of physical observables in the global
gravitational Hilbert space; rather, it also contains non-gauge-invariant
operators, while the only physical operators are those contained 
in  $\agrav$.  It could therefore make sense to only regard 
$\alg$ as an auxiliary structure, in which case it is not important
that the shift symmetry $\theta_s$ does not define a time operator
in $\alg$.  When representing $\agrav$ on a physical Hilbert 
space, one could think of the time operator as living in an enlarged
algebra obtained from $\agrav$ 
by including the generator of $\theta_s$, 
$\wh{\alg} = \agrav \vee \langle \Pi\rangle$.\footnote{This requires
that the commutant of $\agrav$ in the physical representation
is also type $\ttwo_\infty$.  When the commutant is type $\ttwo_1$, 
the trace-scaling automorphism of $\agrav$ is not unitarily implemented.
In fact, trace-scaling automorphisms are the only example
of automorphisms of a factor that may not be unitarily implemented
in some representations \cite{Kadison1955}. }
As discussed above, $\wh{\alg}$ is a type $\tthr_1$ algebra 
isomorphic to the crossed product $\agrav \rtimes_\theta \mathbb{R}$.  
This interpretation has the advantage of giving a possibly more 
general interpretation of how one should characterize observers algebraically:
they are given by a subalgebra inclusion $\agrav\subset \wh{\alg}$, 
where $\wh{\alg}$ is constructed from the gravitational algebra $\agrav$
by including operators associated with the frame data, such as the time
operator.  This interpretation also has the advantage of applying 
in the case that the gravitational algebra is type $\ttwo_1$, 
in which case there cannot be a type $\tthr_1$ quantum field 
theory subalgebra.

\section{Observer in vacuum dS} \label{sec:vacds}

The previous sections focused on the slow-roll inflation example,
where the appearance of a type $\ttwo_\infty$ gravitational 
algebra was a consequence of being the centralizer
of a dominant weight, the Bunch-Davies weight defined in 
(\ref{eqn:psiBD}).  In this section, we will return to the 
original construction of CLPW involving an observer 
in the static patch of dS, and reinterpret the resulting 
algebra again as a centralizer of an integrable weight.  In this
case, the fact that the algebra is type $\ttwo_1$ can be directly
related to the Connes-Takesaki classification of 
integrable weights.  This therefore provides a unifying description
of all current constructions of semifinite gravitational
algebras that also includes crossed product constructions 
\cite{Witten2021, Chandrasekaran2022a, Chandrasekaran2022b,
Jensen2023, Kudler-Flam:2023qfl, Faulkner:2024gst}.
In each case, the gravitational algebra is the centralizer
of an integrable weight, with the type, $\ttwo_1$ or $\ttwo_\infty$,
determined by the properties of the given weight.  

We begin by reviewing the construction of CLPW \cite{Chandrasekaran2022a}, 
which constructs a gravitational algebra for the static
patch of dS via a modification of the crossed product construction.
The starting kinematical algebra consists of an 
algebra $\aqft$ of fields localized to
the static patch.  We take the fields to be ordinary 
matter and graviton fields with potentials that are bounded
below, as opposed to the linear potential model considered
in previous sections.  The modular Hamiltonian of the 
Bunch-Davies weight $\omega_\text{BD}$ generates the boost 
in the static patch, and in this case is expected to act 
ergodically on the quantum fields, so that there 
are no operators in the centralizer $\aqft^{\omega_\text{BD}}$.
To obtain nontrivial boost-invariant operators, one introduces 
an auxiliary Hilbert space $\mathcal{H}_\text{obs} = L^2(\mathbb{R})$ 
associated with an observer, 
where the observer's boost energy $\hobs$ acts as the position
operator on $L^2(\mathbb{R})$, and its canonical conjugate 
is the time operator $\hat{t}$ satisfying $[\hat{t},\hobs]=i$.
Together, the energy and time operators generate 
a type $\tone_\infty$ algebra $\alg_\text{obs}$ of all
operators acting on $\mathcal{H}_\text{obs}$.  

The full boost energy is then given by the sum $H = H_\text{QFT}
+\frac{\hobs}{2\pi}$, where we have normalized the 
observer energy with respect to modular time.  
To see this, we define a weight $\omega_\text{obs}$,
on $\alg_\text{obs}$ with 
density matrix $\rho = e^{-\hobs}$,
\beq
\omega_\text{obs}(\msf{a}_\text{o}) = \tr(e^{-\hobs} \msf{a}_\text{o}),
\quad\msf{a}_\text{o}\in \alg_{\text{obs}}.
\eeq
This weight is a dominant weight on $\alg_\text{obs}$,
and this implies that the tensor product of this weight 
with any faithful weight on $\alg$ is dominant for the total
kinematical algebra $\alg = \alg_\text{QFT}\otimes \alg_\text{obs}$
\cite[Theorem II.1.3(ii)]{Connes1977}.  In particular,
choosing the Bunch-Davies weight for $\alg_\text{QFT}$,
we have that 
\beq \label{eqn:om}
\omega = \omega_\text{BD} \otimes \omega_\text{obs}
\eeq
is dominant, and its modular Hamiltonian is given by
\beq
h = h_0 + \hobs,
\eeq
where $h_0 = 2\pi H_\text{QFT}$ is the modular Hamiltonian
for the quantum fields.

The gauge-invariant algebra consists of operators 
commuting with the boost generator $H$,
and hence it is again the centralizer
of a dominant weight on $\alg$.  In this case, the algebra 
is exactly the crossed product of $\alg_\text{QFT}$ by the 
modular flow generated by $h_0$:
\beq
\alg^\omega = \left\langle e^{-i\hat{t}h_0}\msf{a}
e^{i\hat{t}h_0}, \hobs\right \rangle, \; \msf{a}\in\alg_\text{QFT}.
\eeq
As previously mentioned, this crossed product algebra is 
a type $\ttwo_\infty$ factor.

A final feature of the CLPW construction is to impose that the 
observer energy is positive.  This is done by acting with the 
projection $\msf{e} = \Theta(\hobs)$, where $\Theta$ is the step
function.  Hence the final gravitational algebra is  
given by 
\beq \label{eqn:agrav1}
\agrav = \msf{e} \alg^\omega \msf{e}.
\eeq
The projection $\msf{e}$ is an element of $\alg^\omega$, and 
has a finite trace.  The trace on $\agrav$ is just given by the 
restriction of the trace on $\alg^\omega$, and since $\msf{e}$
becomes the identity operator in $\agrav$, we see that $\agrav$ 
is type $\ttwo_1$.  

This construction of $\agrav$ has an alternative description
directly in terms of a centralizer.  Starting with the 
dominant weight (\ref{eqn:om}), we can first impose
the positive energy projection and construct a non-faithful
weight that is nonzero only on operators with positive
observer energy,
\beq
\omega_{\msf{e}}(\cdot) = \omega(\msf{e} \cdot \msf{e}).
\eeq
In this context, the projection $\msf{e}$ is known as 
the support of the weight $\omega_{\msf{e}}$, being
the largest projection for which $\omega_{\msf{e}}$
defines a faithful weight on the reduced von Neumann
algebra $\msf{e}\alg\msf{e}$.  We can define
the modular flow on this reduced algebra, where it is 
generated by the projected Hamiltonian
\beq
h_{\msf{e}} = h_0 + \hobs \Theta(\hobs).
\eeq
The gravitational algebra $\agrav$ then appears as the 
centralizer of $\msf{e}\alg\msf{e}$, which again reproduces
(\ref{eqn:agrav1}).

The Connes-Takesaki theory of integrable weights can
immediately be applied to conclude that $\omega_{\msf{e}}$ is 
integrable.  To state the classification, we need to 
employ the comparison theory of weights on a von Neumann
algebra \cite[Section XII.4]{TakesakiII}, 
which is somewhat analogous to the 
comparison theory for projections.  Two weights 
$\varphi$ and $\psi$ on a 
von Neumann algebra $\alg$ are called {\it equivalent}
if there exists a partial isometry $\msf{v} \in \alg$
with initial projection $\msf{e} = \msf{v}^*\msf{v}$
equal to the support of $\varphi$ and final projection
$\msf{f} = \msf{vv}^*$ equal to the support of 
$\psi$, such that 
\beq
\varphi(\cdot) = \psi(\msf{v} \cdot \msf{v}^*).
\eeq
This is therefore a generalization of two weights 
being unitarily equivalent that applies to non-faithful 
weights.  We write $\varphi\sim\psi$ when the two 
weights are equivalent.  On the other hand,
if a weight $\varphi$ can be written as 
a projection $\msf{g}$ acting on another weight $\psi$
via $\varphi(\cdot) = \psi(\msf{g} \cdot \msf{g})$, 
$\varphi$ is said to be
a {\it subweight} of $\psi$.  
This defines a partial order on 
weights 
by saying that $\varphi\precsim \psi$ if $\varphi$ is 
equivalent to a subweight of $\psi$, meaning there 
exists a partial isometry $\msf{v}$ and a 
projection $\msf{g}$ such that 
\beq
\varphi(\cdot) = \psi(\msf{gv}\cdot\msf{v}^*\msf{g})
\eeq
The classification theorem \cite[Theorem II.2.2]{Connes1977}\cite[Theorem XII.4.21]{TakesakiII}
then states that a weight $\varphi$
is
integrable if and only if $\varphi\precsim \omega$, 
where $\omega$ is a dominant weight.  Hence, because
$\omega_\msf{e}$ defined above
is a subweight of the dominant weight $\omega$, we immediately
conclude that it is integrable.  

Although in this example $\omega_{\msf{e}}$ is not a 
faithful weight on $\alg$, there are cases where a 
weight is faithful and integrable, but at the same
time not dominant,
so that it possesses a 
type $\ttwo_1$ centralizer. In fact, the weight
defined above is one such example if we take $\omega_{\msf{e}}$ 
to be defined on the reduced algebra $\alg_{\msf{e}} = 
\msf{e}\alg\msf{e}$.  This reduced algebra is still
a type $\tthr_1$ factor, and, by definition, 
$\omega_{\msf{e}}$ is faithful on it.  However, we cannot
say that $\omega_{\msf{e}}$ is a subweight of $\omega$ 
on the reduced algebra, since $\omega$ is a weight 
on a different algebra $\alg$.  This is where 
it is important to use the comparison theory for weights,
since we only need to find a weight equivalent to 
$\omega_{\msf{e}}$ that is a subweight of a dominant 
weight on $\alg_{\msf{e}}$.  Hence, we need 
to find a dominant weight $\omega_D$ on $\alg_{\msf{e}}$ and 
a partial isometry $\msf{v} \in \alg_{\msf{e}}$ and 
projection $\msf{g}\in \alg_{\msf{e}}$
such that
\beq \label{eqn:wesubweight}
\omega_{\msf{e}} = \omega_{D}(\msf{gv}\cdot \msf{v}^*\msf{g}).
\eeq

To find the dominant weight on $\alg_{\msf{e}}$, we note 
that because $\alg$ is type $\tthr$, $\msf{e}$ is 
an infinite projection, and hence is equivalent
to the identity \cite[Proposition V.1.39]{TakesakiI}.  
There then exists an isometry $\msf{u}\in\alg$ satisfying
\beq \label{eqn:e}
\msf{u}^*\msf{u} = \mathbbm{1},\quad \msf{uu}^* = \msf{e}.
\eeq
Conjugating $\alg$ by $\msf{u}$ exhibits the isomorphism
between $\alg$ and $\alg_{\msf{e}}$, i.e.\ $\msf{u}\alg
\msf{u}^* =\alg_{\msf{e}}$.  We can then 
define a dominant weight $\omega_D$ on $\alg_\msf{e}$ 
as the equivalent weight to $\omega$ under the 
co-isometry $\msf{u}^*$,
\beq
\omega_D(\cdot) = \omega(\msf{u}^* \cdot \msf{u}).
\eeq
The support projection of $\omega_D$ is $\msf{e}$, and 
hence it is faithful on $\alg_{\msf{e}}$, and we 
can check that it possesses 
the defining property of a dominant weight,
namely being unitarily related to the rescaled 
weight $\lambda \omega_D$ for any $\lambda>0$
\cite[Theorem XII.4.18]{TakesakiII}. This readily follows
from the fact that $\omega$ is dominant on $\alg$, so 
for any $\lambda>0$, there is a unitary 
operator $\msf{w}\in\alg$ such that $\omega(\msf{w}\cdot
\msf{w}^*) = \lambda\omega(\cdot)$.  Then
the operator $\msf{uwu}^*$ performs the same 
function on $\omega_D$:  using (\ref{eqn:e}), we find
\beq
\omega_D(\msf{uwu}^* \cdot \msf{uw}^*\msf{u}^*)
=
\omega(\msf{u}^*\msf{uwu}^* \cdot\msf{uw}^*\msf{u}^*\msf{u})
=
\lambda\omega(\msf{u}^* \cdot \msf{u})
=
\lambda\omega_D(\cdot).
\eeq

Now we consider the image of $\msf{u}$ under the isomorphism,
\beq \label{eqn:v}
\msf{v} = \msf{uuu}^* = \msf{eue}.
\eeq
This is a partial isometry in $\alg_\msf{e}$ with 
initial and final projections 
\beq
\msf{e} = \msf{v}^* \msf{v},\quad \msf{g} = \msf{vv}^* = 
\msf{ueu}^*.
\eeq
The claim is that $\msf{v}$ maps $\omega_\msf{e}$ to the 
subweight 
 $\omega_D(\msf{g}\cdot \msf{g})$ of the dominant
weight $\omega_D$.  We  compute
\begin{align}
\omega_D(\msf{g}\cdot \msf{g}) 
&=
\omega(\msf{u}^* \msf{g} \cdot \msf{gu})
=
\omega(\msf{u}^*\msf{ueu}^*\cdot\msf{ueu}^*\msf{u})
=
\omega(\msf{e u}^* \cdot \msf{ue})
=
\omega(\msf{v}^*\cdot \msf{v})
=
\omega(\msf{ev}^*\cdot \msf{ve})
\nonumber \\
&=
\omega_{\msf{e}}(\msf{v}^* \cdot \msf{v}).
\end{align}
This relation then shows that 
\beq
\omega_{\msf{e}}(\cdot) = \omega_{\msf{e}}(\msf{v}^*\msf{v}\cdot
\msf{v}^*\msf{v}) = \omega_D(\msf{gv}\cdot \msf{v}^*\msf{g}),
\eeq
verifying (\ref{eqn:wesubweight}).  Hence, $\omega_{\msf{e}}$ 
is equivalent to a subweight of a dominant weight on $\alg_{\msf{e}}$,
as is required for it to be an integrable weight.

This discussion shows that quite generally, the semifinite
algebras that have appeared in recent works on semiclassical
gravity all have a description as a centralizer of 
an integrable weight.  One difference when the 
algebra is type $\ttwo_1$ is that there cannot be a type
$\tthr_1$ subalgebra associated with the pure
quantum field degrees of freedom 
with which to construct a crossed product
description.  This follows from the basic fact that all
subalgebras of a type $\ttwo_1$ algebra have a tracial
state defined on them, coming from the restriction
of the canonical trace on the original algebra.  
Hence there can be no infinite subalgebras,
so in particular no factors of type $\tone_\infty$, $\ttwo_\infty$,
or $\tthr$.  The existence of a crossed product 
description in the inflationary example 
was related in section \ref{sec:crossprod} 
to the separation on the gravitational 
algebra into quantum field degrees of freedom and an 
observer.  Hence, it is less clear how to make this 
separation when the centralizer is type $\ttwo_1$.

One comment on this is that there still exist operators in
$\alg_\msf{e}$ that play the role of time operators, in the sense
that they are eigenoperators of $\sigma_t$, 
the modular flow of the 
integrable weight.  If $\msf{a}\in\alg_\msf{e}$ is 
an operator for which the modular time-average (\ref{eqn:Tovw}) 
converges, then the following Fourier transform
of this operator with respect to modular 
time is also convergent \cite[Lemma II.2.3]{Connes1977},
\beq
\msf{a}_\omega = \int_{-\infty}^\infty dt \sigma_t(\msf{a})
e^{-i\omega t}.
\eeq
It is straightforward to verify that $\sigma_u(\msf{a}_\omega)
= e^{i\omega u} \msf{a}_\omega$, showing that it an
eigenoperator of the modular flow.  Performing a polar
decomposition on the eigenoperator $\msf{a}_\omega
= \msf{w}_\omega |\msf{a}_\omega|$, we see that the 
partial isometry $\msf{w}_\omega$ is the modular
flow eigenoperator
with eigenvalue $e^{i\omega u}$ and the norm $|\msf{a}_\omega|$
is in the centralizer.  It is also clear that $\msf{w}_\omega$
maps the centralizer $\agrav$ to itself, since if $\msf{b}\in
\agrav$, we  have
\beq
\sigma_u(\msf{w}_\omega \msf{b} \msf{w}_\omega^*)
=
e^{i\omega u} \msf{w}_\omega \msf{b} \msf{w}_\omega^* e^{-i\omega u}
=
\msf{w}_\omega \msf{b} \msf{w}_\omega^*.
\eeq
Hence, $\msf{w}_\omega$ behaves like the exponentiated time
operator $e^{i\omega\hat{t}}$.  However, it does not 
define an automorphism of $\agrav$, since in general 
$\msf{w}_\omega$ is not unitary.  This is important, since 
otherwise we would find a trace-scaling automorphism 
of the $\ttwo_1$ factor $\agrav$, but this cannot 
occur since all automorphisms of type $\ttwo_1$ factors 
are trace-preserving.

These time operators have a  more explicit 
description in terms of the dominant weight $\omega_D$ 
defined 
on $\alg_\msf{e}$.  There, the time operators are given by 
$\msf{u} e^{i\omega{\hat{t}}} \msf{u}^*$, where $\hat{t}$ 
is the original time operator defined on $\alg$ in the 
discussion of the crossed product, and $\msf{u}$ 
is defined above (\ref{eqn:e}). We then map this 
to a time operator for the faithful
integrable weight $\omega_\msf{e}$ using the partial isometry
$\msf{v}$ defined in (\ref{eqn:v}).  This results in the 
operators $\msf{v}^* \msf{u} e^{i\omega\hat{t}} \msf{vu}^*$
which are then the eigenoperators for the modular 
flow of $\omega_e$.  From (\ref{eqn:e}) and (\ref{eqn:v}), 
this operator is equivalent to $\msf{e}e^{i\omega\hat{t}}\msf{e}$,
which is just the projection of the time operator in the 
larger algebra $\msf{a}$ to the support of $\omega_\msf{e}$.  

Hence, a somewhat general picture emerges for how to view
the time operators.  One could characterize them as a subfactor
inclusion $\agrav\subset \aext$, where $\agrav$ is a semifinite
gravitational algebra possessing a trace, and $\aext$ is an
extended algebra that includes time operators.  The interesting
case in our context seems to be when $\aext$ is a type $\tthr_1$
factor associated with quantum fields, and the inclusion
is irreducible, meaning that the relative commutant
$\agrav'\wedge \aext$ is trivial, consisting of operators 
proportional to the identity.  Semifiniteness 
of $\agrav$ means that there must be an 
operator-valued weight $\Tovw: \aext\rightarrow\agrav$
\cite{Haagerup1979II},
which, as discussed above, always occurs when
$\agrav$ is the centralizer of an integrable weight.  
The time operators are then the additional operators one
must add to $\agrav$ to construct the full
extended algebra $\aext$.  For the case of a general subfactor
inclusion, these additional operators are related to the 
Pimsner-Popa basis for the inclusion \cite{Pimsner1986,Herman1989,
Nill1995, Enock1996}.

Although the subfactor picture gives a more general definition 
of time operators, it does not give a proposal for the 
observer Hamiltonian $\hobs$ in the general case.  One such
proposal valid in the case of a centralizer of an integrable
weight is to use the map $\msf{v}$ to find an equivalent 
weight that is a subweight of a dominant weight $\omega_D$.  
The centralizer of the dominant weight admits a crossed-product
description, which then results in an observer Hamiltonian, subject
to choices of frame, as described in section \ref{sec:crossprod}.
We can then map these observer Hamiltonians back to the 
original algebra using $\msf{v}^*$ to obtain a collection
of operators in the type $\ttwo_1$ algebra that 
behave like the projected observer Hamiltonian 
$\hobs\Theta(\hobs)$.  This then gives an algebraic way
of characterizing observers in situations where a 
direct crossed product description is 
not available.

\section{Discussion}
\label{sec:discussion}

This work has sought to provide a unifying explanation
for the occurrence of 
type $\ttwo$ algebras in semiclassical gravity and to
clarify the role of observers in their construction.  
This was done by way of  example through the slow-roll 
inflation model considered by Chen and Penington \cite{Chen2024}.
The gravitational algebra
in this case is the centralizer of the 
Bunch-Davies weight, and is nontrivial due to the linear
potential chosen for the scalar field.  We emphasized that the 
appearance of a trace and the associated renormalized 
entropy is a consequence of being the centralizer of a modular
flow, and we proposed that this is the crucial feature 
shared by all recent examples of semifinite
gravitational algebras.  We also gave a canonical description
of the inflationary algebra as a crossed product
by identifying a preferred trace-scaling automorphism.  
This allowed for the identification of an observer 
degree of freedom constructed intrinsically from the 
quantum field degrees of freedom.  The ambiguity in the 
choice of the observer Hamiltonian was related to a kind
of quantum reference frame dependence of the algebra's description.
From this, we see that the existence of an observer
is an additional structure attached to the algebra
needed to represent it as a crossed product, and 
is not in itself directly responsible for the 
existence of a trace.  We also discussed a more general
definition of observer degrees of freedom that 
would be valid in the type $\ttwo_1$ case in which no
modular crossed product description is permitted.  We suggested
that the appropriate way to understand observers in this 
context is via a subfactor inclusion $\agrav\subset\alg$,
where the larger algebra $\alg$ is  a  physical
algebra that includes time operators for the observer in 
$\agrav$.  

We conclude with a few comments on some interesting 
directions for future investigations.

\subsection{Centralizers and semiclassical gravity}

A clear outcome of this investigation is that 
centralizers of weights are potentially very interesting
objects to study in the context of semiclassical gravity
(this point has also been advocated in \cite{Gomez:2024kuy,
Gomez:2023jbg, Gomez:2023tkr}).  
In the present work, we focused specifically on 
integrable weights, which are those for which the 
modular time-averaging operation (\ref{eqn:Tovw}) is useful,
in that it defines a semifinite operator-valued weight.  
This characterization could  potentially
 be used beyond the free field theory examples considered here.
A natural question to consider is what sorts of potentials
give rise to integrable Bunch-Davies weights, such as the one
found in the present work.  The appearance of a type $\ttwo_\infty$
centralizer was clearly related to the apparent instability of the model, with the potential for $\phi$ being unbounded below.  
It would be interesting to determine whether this is the only
feature necessary, and whether more general interacting theories
admit a natural Bunch-Davies weight that is integrable.  

A related question concerns  when type $\ttwo_1$ centralizers
can arise.  If we make the scalar field potential
bounded below by giving it a mass, we expect the
Bunch-Davies wavefunction to be a normalizable state
with an ergodic modular flow, and hence not to have 
an interesting centralizer.  One can then ask if there 
is any natural potential that leads to a type $\ttwo_1$
centralizer, which we can phrase mathematically as whether
the Bunch-Davies weight can ever be faithful, integrable, and 
non-dominant.  A similar question can be posed when 
including an explicit observer in the static patch of de
Sitter, as in the CLPW construction.  There, the 
type $\ttwo_1$ algebra was obtained by explicitly 
projecting to a finite subalgebra, but we could also 
ask whether a different choice of observer Hamiltonian 
still leads to an integrable weight but does not
involve the explicit projection.  A related question
in this case is whether interactions between the observer
and quantum fields can destroy integrability of the weight. 
The characterization of the allowed interactions preserving
integrability likely has an interesting connection to the 
cohomology discussed 
in section \ref{sec:crossprod} of the Connes cocycles.  

While we focused in this work on integrable weights, these 
are not the only weights on a type $\tthr_1$ algebra 
with nontrivial centralizer.  Another interesting 
class of weights are the strictly semifinite weights mentioned 
in section \ref{sec:intweight}.  These weights also
have a modular-time-averaging procedure, but it involves 
a normalized average \cite{Haagerup1979II}:
\beq
\mathcal{E}(\msf{a}) = \lim_{\Lambda\rightarrow\infty}
\frac{1}{2\Lambda} \int_{-\Lambda}^\Lambda dt\sigma_t(\msf{a}).
\eeq
In this case, $\mathcal{E}$ defines a conditional
expectation, i.e.\ a normalized operator-valued weight
satisfying $\mathcal{E}\circ\mathcal{E} = \mathcal{E}$.  
Note that the hyperfinite type $\tthr_1$ algebra $\rinj_\infty$
prevalent in quantum field theory admits many
such weights with type $\ttwo$ factors as centralizers
\cite{Haagerup1987}, hence
it would be interesting to see if they have any applications
to the current gravitational algebra constructions.  
An interesting comment in 
the case that the centralizer has trivial relative
commutant in the type $\tthr_1$ algebra is that  the modular
operator $\Delta$ for such weights is diagonalizable,
meaning there exists a complete basis of normalizable 
eigenstates for it \cite{Connes1974, Penneys2019}.\footnote{Since $\rinj_\infty$ it 
type $\tthr_1$, the eigenvalues associated with this 
basis must form a dense subset of $\mathbb{R}^+$.} 
Such a weight is known as {\it almost periodic}.  
If there is a physically relevant model involving such 
a weight, it would be interesting to determine the properties
of the time operator in this case.  As the conjugate of 
an operator with a discrete spectrum, the time operator
ought to have an associated periodicity.  One might 
hope to be able to characterize this time operator
using the subfactor description 
of a type $\ttwo_1$ algebra embedded in a type $\tthr_1$ algebra,
perhaps using the techniques developed in \cite{Penneys2019}.

\subsection{Quantum reference frames} \label{sec:qrf}

One of the outcomes of representing the inflationary gravitational
algebra as a crossed product in 
section \ref{sec:intweight} is that it lead to the 
identification of a class of observer Hamiltonians, each
of which was associated with a choice of quantum reference frame
for the quantum field subalgebra.  This then ties 
the present work to a number of others that have 
emphasized the connection between crossed products
and quantum reference frames 
\cite{Fewster:2024pur, DeVuyst:2024pop,
AliAhmad:2024wja, DeVuyst:2024uvd}.  The identification
of the observer Hamiltonian intrinsically from the 
quantum fields is a distinguishing feature of the present
work, but is closely related to the top-down approach
to quantum reference frames discussed in \cite{AliAhmad:2024wja}.
An important consequence of this intrinsic observer is that the
entropy computed for states on the algebra is frame-independent: 
the choice of observer simply gives a way to decompose 
the algebra as a crossed product, but the entropy only depends on 
the state of the algebra as a whole, and is thus independent
of this choice of frame.  This contrasts with the conclusions
found in \cite{DeVuyst:2024pop,
 DeVuyst:2024uvd}, where the entropy was argued to be 
frame-dependent. This difference stems from the frame being 
intrinsically defined in the present work, as opposed to being
externally imposed.  

We also discussed in section \ref{sec:vacds} how one 
should interpret the time operator and the quantum
reference frame description in the type $\ttwo_1$ case
where the crossed product is not applicable.  It would be good
to compare these ideas to those presented in 
\cite{Fewster:2024pur, DeVuyst:2024pop, DeVuyst:2024uvd}.
In particular, \cite{DeVuyst:2024pop, DeVuyst:2024uvd}
interpreted the type $\ttwo_1$ algebras as situations
where one has a non-ideal clock, and it would be useful 
to spell out this characterization in the present model.
Additionally, the Connes-Takesaki classification shows 
that since the type $\ttwo_1$ examples involving 
an observer in de Sitter space arise as centralizers 
of integrable weights, there is always a description where 
the algebra appears as a finite projection acting
on a crossed product algebra.  This description may 
provide a way to characterize non-ideal clocks more
broadly.  

One of the key questions addressed in the work 
\cite{Fewster:2024pur} was the characterization
of when the gravitational algebra is semifinite.  They
gave a sufficient condition for this to occur, which 
required that one form a crossed product by a group containing
the modular automorphism group as a factor.  They did not,
however, identify whether this condition is necessary.  
A stronger result was derived in \cite{AliAhmad:2024eun},
where it was shown that if one takes a crossed product 
by any group containing the modular automorphism group,
the resulting algebra is semifinite only if 
the modular automorphism group is central.  This result 
is perhaps not so surprising in light of the fact that
modular flow is always central in the outer automorphism group
(see the discussion in appendix \ref{sec:automorphism}).
Hence, up to inner automorphisms, modular flow always appears
as a central generator in a given crossed product 
construction.  A stronger result proved in \cite{Jensen:2024dnl} 
showed that the only crossed product of a type $\tthr_1$ factor
that leads to a semifinite algebra is the modular crossed 
product, thereby demonstrating that this is a necessary 
condition.  In the present work, we have emphasized that 
gauging modular flow is the key aspect leading to 
a semifinite algebra.  In the most general case,
the gravitational algebra appears as a subfactor
$\agrav\subset \alg$, with $\agrav$ the centralizer
of a weight on $\alg$.  A theorem by Haagerup \cite[Theorem
5.7]{Haagerup1979II}
then demonstrates that semifiniteness of $\agrav$ is 
equivalent to the existence of an operator-valued weight
$\Tovw:\agrav\rightarrow\alg$.  This description
in terms of subfactors and operator-valued weights 
appears to be the most general case possible, and it would 
be worth considering whether other mathematical results 
involving subfactor theory could be useful
in understanding gravitational algebras
\cite{Jones1983, Herman1989, Nill1995, Enock1996}.  
See \cite{vanderHeijden:2024tdk} 
for some recent applications of subfactor 
theory to the physics of black holes.

\subsection*{Acknowledgments}

I thank Tom Faulkner, Ben Freivogel, Oliver Janssen, 
Victor Gorbenko, David Penneys, Geoff Penington, 
and Kamran Vaziri for 
helpful discussions. This 
research is supported by  the Air Force Office of Scientific Research under award number FA9550-19-1-036, and from the Heising-Simons Foundation ‘Observational Signatures of
Quantum Gravity’ QuRIOS collaboration grant 2021-2817.

\appendix

\section{Automorphism groups of hyperfinite factors}
\label{sec:automorphism}

We want to take advantage of some classification results
of hyperfinite factors in order to conclude certain properties
about the structure of the gravitational algebra.  Here we will
give a brief summary of these classification theorems, and provide
some comments on the structure of automorphism groups for 
more general factors.  We will restrict attention to 
separable von Neumann algebras below, meaning they always 
have a representation on a separable Hilbert space.  See 
\cite{Stormer1993, Masuda2017} for fairly concise overviews, 
and \cite[Chapter 5]{Connes1994}, \cite[Chapters
XIV, XVII, XVIII]{TakesakiIII} for more detailed 
treatments.

Given a von Neumann algebra $\algm$, its automorphism
group $\aut(\algm)$ is the set all bijective maps 
$\alpha:\algm\rightarrow\algm$ such that $\alpha(\msf{a}\msf{b}) = 
\alpha(\msf{a})\alpha(\msf b)$ and $\alpha(\msf{a}^*) = 
\alpha(\msf{a})^*$.  There are a number of  normal 
subgroups of $\aut(\algm)$ that arise in its 
characterization.  The first of these is the group
$\Int(\alg)$ of {\it inner automorphisms}, generated by
the adjoint actions $\Ad_{\msf{u}}  = \msf{u}
(\cdot) \msf{u}^*$ by unitary operators $\msf{u}\in\algm$.
The {\it outer automorphism group} $\out(\algm)$ is the 
quotient group $\aut(\algm)/\Int(\algm)$, consisting  of 
equivalence classes of automorphisms modulo inner
automorphisms, $\alpha\sim \Ad_{\msf u}\circ\,\alpha$.  
For type $\tone$ factors, all automorphisms are inner 
\cite[Lemma XI.3.7]{TakesakiII}, so the 
only factors with nontrivial outer automorphism groups 
are type $\ttwo$ or type $\tthr$
(although there also 
exist some type $\ttwo$ factors for which all
automorphisms are inner \cite{Ioana2008}).  

The next normal subgroup is the group of {\it approximately inner
automorphisms}, denoted $\overline{\Int(\algm)}$.  This group
consists of automorphisms that arise as limits of inner automorphisms,
so in particular $\Int(\algm)\subset\overline{\Int(\algm)}$.  
These limits are taken with respect to a topology on $\aut(\algm)$
inherited from the strong operator topology on the unitary
operators implementing the automorphisms in a standard form
representation of $\algm$ \cite{Araki1974, Haagerup1975}
\cite[Section IX.1]{TakesakiII}.  In this standard 
implementation, the Hilbert space $\hs$ is  a 
GNS representation of $\algm$ with respect to a faithful 
normal state (or more generally, a semicyclic representation
with respect to a faithful semifinite normal weight 
\cite[Section VII.1]{TakesakiII}), and each automorphism $\alpha$
maps to a unique unitary operator $U_\alpha \in 
\mathcal{B}(\hs)$ that commutes with the conjugation $J U_\alpha J
 = U_\alpha$ and preserves the natural cone $U_\alpha 
 \mathcal{P}^\natural = \mathcal{P}^\natural$ associated 
 to the representation.  A sequence of automorphisms $\alpha_n$ 
limits to $\alpha$ in this topology if the corresponding 
standard unitaries $U_{\alpha_n}$ limit to $U_\alpha$ strongly,
meaning that 
\beq
\big\lVert(U_{\alpha_n} - U_\alpha)|\psi\rangle\big\rVert
\overset{n\rightarrow \infty}{\longrightarrow} 0,\qquad
\forall\: |\psi\rangle \in \hs.
\eeq
Note that for an inner automorphism $\beta\in\Int(\algm)$, 
the implementing unitary factorizes according to 
$U_\beta = \msf{u}_\beta \msf{u}_\beta'$, with $\msf{u}_\beta\in
\algm$ and $\msf{u}_\beta' \in \algm'$.  This factorization
fails for an approximately inner automorphism $\alpha
\in \overline{\Int(\algm)}$ that is not inner,
since although $U_\alpha$ is a limit of factorizing unitaries 
$U_{\alpha_n} = \msf{u}_{\alpha_n} \msf{u}_{\alpha_n}'$, 
the individual factors $\msf{u}_{\alpha_n}$, $\msf{u}_{\alpha_n}'$ 
 fail to have well-defined limits. 
 
This sequence of approximate factorizations that exists for 
approximately inner automorphisms appears to be the reason
that one can think of the modular automorphism 
of hyperfinite type $\tthr$ factors as being 
generated by singular density matrices.  Occasionally 
it is helpful to employ
the formal expression $\Delta^{is} = \rho^{is}(\rho')^{-is}$
to denote the generator of modular flow on an algebra,
where $\rho$ and $\rho'$ are the density matrices
for the state on the algebra  and its commutant.  This expression
is valid for type $\tone$ and $\ttwo$ algebras which admit 
well-defined density matrices, but is not correct in the 
type $\tthr$ case since the density matrix is not defined.  
However, one could approximate these density
matrices using a sequence $U_n(s) = 
\msf{u}_n(s) \msf{u}_n'(-s)$ that limits to the 
modular automorphism generator $\Delta^{is}$.  Then
we could define the regulated density matrix by the relation
$\rho_n^{is} = e^{-isK_n} = \msf{u}_n(s)$.  This leads to a 
proposal for defining the regulated {\it entanglement Hamiltonian}
(sometimes referred to as the one-sided modular Hamiltonian)
$K_n = -\log\rho_n$.  Although $K_n$ cannot limit to a well-defined
unbounded operator as $n\rightarrow\infty$ due to the 
fact that modular flow is outer for almost all values of $s$ 
on type $\tthr$ factors, we expect that $K_n$ will limit to 
a  sesquilinear form $K$, which has finite 
expectation values in a dense set of states.  In this case, 
we conjecture that the modular Hamiltonian $h = -\log \Delta$ 
factorizes into entanglement Hamiltonians $h = K - K'$,
where $K$ and $K'$ are sesquilinear forms, whenever
modular flow is approximately inner.  

This factorization
of $h$ was a crucial feature used in \cite{Kudler-Flam:2023hkl}
to give an invariant definition of entanglement entropy differences
for type $\tthr_1$ factors appearing in quantum
field theory.  Interestingly, there exist type $\tthr$ factors
in which the modular automorphism is not approximately
inner; for example, there are {\it full factors} in which
all approximately inner automorphisms are actually inner,
and hence do not include the modular automorphism 
\cite{Connes1974}.  For such algebras, the definition
of entanglement entropy differences proposed 
in \cite{Kudler-Flam:2023hkl} would not work.  As we discuss
below, modular flow is approximately inner for the 
hyperfinite $\tthr_1$ factor $\rinj_\infty$, which 
suggests that the ability to compute entropy differences
in quantum field theory is closely tied to hyperfiniteness
of the local algebras.  It would be interesting to investigate
this point further, and to understand if an alternative notion
of entropy differences exists for (non-hyperfinite) algebras in 
which modular flow is not approximately inner.

For the hyperfinite $\ttwo_1$ factor $\rinj_0$, it turns out that all of its 
automorphisms are approximately inner, $\aut(\rinj_0) = \overline{\Int(\rinj_0)}$
\cite[Theorem XIV.2.16]{TakesakiIII}. In this case, the inner automorphisms 
are a maximal normal subgroup
of $\aut(\rinj_0)$, implying that $\out(\rinj_0)$ is a simple group
\cite{Connes1975}\cite[Corollary XVII.3.21]{TakesakiIII}.  
The only other  hyperfinite factor for which all automorphisms are 
approximately inner is the hyperfinite $\tthr_1$ factor $\rinj_\infty$,
so in this case $\aut(\rinj_\infty) = \overline{\Int(\rinj_\infty)}$ as well
\cite{Connes1985}\cite[Theorem XVIII.4.29]{TakesakiIII}.  However,
we will see shortly that $\out(\rinj_\infty)$ is not simple,
and in fact has a center coinciding with the modular automorphism
group.  

This last point requires the introduction of the normal subgroup
$\cnt(\algm)$ of {\it centrally trivial automorphisms}. These
automorphisms are defined in terms of their action on {\it 
strongly central sequences}, which are 
bounded sequences of operators $\msf{x}_n\in\algm$ that 
asymptotically commute with all linear functionals 
in the predual $\algm_*$.  More precisely, for any $\omega\in
\algm_*$, we can define new linear functionals $\msf{x}_n\omega$
and $\omega \msf{x}_n$ by 
\beq
\msf{x}_n\omega (\msf{a}) = \omega(\msf{a} \msf{x}_n), \qquad
\omega\msf{x}_n(\msf{a}) = \omega(\msf{x}_n \msf{a}).
\eeq
Then the sequence $(\msf{x}_n)$ is strongly central if 
\beq
0= \lim_{n\rightarrow\infty} \lVert 
\omega\msf{x}_n -\msf{x}_n\omega \rVert
=
\lim_{n\rightarrow\infty} 
\sup_{
\substack{
{\msf{a}\in\algm}\\
{\lVert\msf{a}\rVert\leq 1}
}
}
\Big|\omega\big([\msf{x}_n,\msf{a}]\big)\Big|
\quad \forall \omega \in \algm_*.
\eeq
Two strongly central sequences $(\msf{x}_n)$ and 
$(\msf{y}_n)$ are said to be equivalent if the difference
$\msf{x}_n - \msf{y}_n$ converges to $0$ in the $\sigma$-strong$^*$
topology, and a sequence $(\msf{x}_n)$ is called trivial if it is 
equivalent to a sequence $(\msf{a}_n)$ in which all $\msf{a}_n$ are 
in the center of $\algm$.  Nontrivial strongly central
sequences exist for any algebra in which $\overline{\Int(\algm)}
\neq \Int(\algm)$ \cite{Connes1974}\cite[Theorem XIV.3.8]{TakesakiIII} (i.e.\ whenever $\algm$ is not full),
so in particular all hyperfinite 
algebras admit such nontrivial sequences.  An automorphism $\alpha\in\aut(\algm)$
then is called centrally trivial if its action on every strongly central
sequence yields an equivalent sequence, i.e.\ $\alpha(\msf{x}_n) - 
\msf{x}_n$ converges $\sigma$-strong$^*$-ly to $0$ for every
central sequence $(\msf{x}_n)$.  

All inner automorphisms are centrally trivial since $\msf{u}\msf{x}_n
\msf{u}^* - \msf{x}_n = \msf{u}[\msf{x}_n,\msf{u^*}]$, and 
$[\msf{x}_n,\msf{u}^*]$ converges to zero $\sigma$-strong$^*$-ly
whenever $\msf{x}_n$ is strongly central 
\cite[Lemma XIV.3.4]{TakesakiIII}.  Hence it is always the 
case that $\Int(\algm)\subset
\cnt(\algm)$.  For the hyperfinite type $\ttwo_1$ factor $\rinj_0$ 
and $\ttwo_\infty$ factor 
$\rinj_{0,1} = \rinj_0\otimes \mc{F}_\infty$ (where $\mc{F}_\infty$
is the unique type $\tone_\infty$ factor of all bounded operators
on an infinite separable Hilbert space), all centrally trivial 
automorphisms are inner, so $\cnt(\rinj_0) = \Int(\rinj_0)$, 
$\cnt(\rinj_{0,1}) = \Int(\rinj_{0,1})$ \cite{Connes1975}
\cite[Theorem XIV.4.16, Lemma XVII.3.11]{TakesakiIII}.  
On the other hand, these equalities do not hold for type 
$\tthr$ factors, since one can show that modular automorphisms
$\sigma_t^\varphi$ are always centrally trivial 
\cite{Connes1974}\cite[Proposition XVII.2.12]{TakesakiIII}, 
but in any type $\tthr$ algebra, modular 
automorphisms are not inner for almost all values of $t$
\cite{Connes1973}.  

Centrally trivial and approximately inner automorphisms commute
with each other up to elements of $\Int(\algm)$; i.e.\
$\cnt(\algm)/\Int(\algm)$ and $\overline{\Int(\algm)}/
\Int(\algm)$ are commuting subgroups of $\out(\algm)$
\cite{Connes1975}\cite[Lemma XIV.4.14]{TakesakiIII}.  For hyperfinite factors, $\cnt(\algm)/
\Int(\algm)$ is the centralizer of $\overline{\Int(\algm)}/
\Int(\algm)$, so it contains all automorphisms that outer-commute
with the approximately inner 
automorphisms \cite{Connes1975}\cite[Corollary XVII.2.11]{TakesakiIII}.\footnote{In fact this occurs
whenever $\algm$ is isomorphic to $\algm\otimes \rinj_0$;
such factors are called {\it strongly stable} \cite[Definition
XIV.4.1]{TakesakiIII}.  }  Since all automorphisms of the 
hyperfinite $\tthr_1$ factor $\rinj_\infty$ are approximately
inner, this shows that $\cnt(\rinj_\infty)/\Int(\rinj_\infty)$
is the center of $\out(\rinj_\infty)$.  One can further 
show that any automorphism in $\cnt(\rinj_\infty)$ 
is a combination of a modular automorphism and an inner
automorphism 
\cite{Kawahigashi1992}\cite[Theorem XVIII.4.29]{TakesakiIII}, 
which then implies that $\cnt(\rinj_\infty)/
\Int(\rinj_\infty)$ coincides with the image of modular
automorphisms in $\out(\rinj_\infty)$.  Hence,
modular flows define the center of the outer
automorphism group $\out(\rinj_\infty)$.  

There is a final normal subgroup that is relevant for 
type $\ttwo_\infty$ and type $\tthr_\lambda$ algebras 
with $\lambda\neq 1$, 
identified in \cite{Haagerup1990b} as the collection
of {\it approximately pointwise inner automorphisms}.  
These are automorphisms $\alpha$ that can be approximated by
inner automorphisms in a 
state-dependent manner, meaning that given a 
state $\varphi$ and an error tolerance $\vep$, one 
can find a unitary $\msf{u} = \msf{u}(\varphi,\vep)\in
\algm$ such that $\varphi\circ \alpha^{-1}$ and $\msf{u}\varphi
\msf{u}^{-1}$ are close in norm, i.e.
\beq
\lVert \varphi\circ\alpha^{-1} - \msf{u}\varphi\msf{u}^{-1}
\rVert <\vep.
\eeq
This group of automorphisms will be denoted $\uni(\algm)$.  
These automorphisms can equivalently be described as 
the kernel
of the Connes-Takesaki $\Mod$ homomorphism \cite{Connes1977},
which we now review.  

The $\Mod$ homomorphism is easiest to describe
for $\ttwo_\infty$ factors.  In that case there is a 
tracial weight $\tau$ that is unique up to rescaling, and 
hence any automorphism $\alpha$ must preserve the trace up to 
a rescaling: $\tau\circ\alpha = e^{-s} \tau$, $e^{-s}\in
\mathbb{R}_+$.  The number $e^{-s}$ is known as the module
of the automorphism, and the map $\Mod(\alpha) = e^{-s}$
is a homomorphism from $\aut(\algm)$ into $\mathbb{R}_+$.  
$\uni(\algm)$ is the kernel of this homomorphism, so
in the type $\ttwo_\infty$ case it coincides with the 
trace-preserving automorphisms.
There is also
a definition of the $\Mod$ homomorphism for 
type $\tthr$ algebras, arising from the fact that 
any automorphism $\alpha$ induces an automorphism
on the smooth flow of weights, which is the center
of the modular crossed product algebra \cite{Connes1977}\cite[Section XII.4]{TakesakiII}.  
Note that $\Mod$ is the trivial homomorphism
for type $\ttwo_1$ and type $\tthr_1$ algebras, and so
in these cases $\uni(\algm) = \aut(\algm)$.  This occurs
because all automorphisms of a $\ttwo_1$ factor are 
trace-preserving, and because the flow of weights is 
trivial for a $\tthr_1$ factor.
It is straightforward to show that $\Int(\algm) \subset
\uni(\algm)$, and because the $\Mod$ homomorphism is continuous,
it follows that $\overline{\Int(\algm)} \subset \uni(\algm)$
\cite{Connes1977}.  For any hyperfinite algebra $\rinj$, 
this last inclusion is saturated, so we have that 
$\uni(\rinj) = \overline{\Int(\rinj)}$ 
\cite{Kawahigashi1992}.
In general, $\cnt(\algm)$ need not 
be a subgroup of $\uni(\algm)$, but for a hyperfinite 
factor $\rinj$ 
the inclusion $\cnt(\rinj)\subset\uni(\rinj)$ does hold
\cite{Haagerup1990, Kawahigashi1992}.  

We will be interested in the trace-scaling automorphisms
for the hyperfinite $\ttwo_\infty$ factor $\rinj_{0,1}$.  
In this case,
there is a unique automorphism, up to conjugation, 
for any given value of the module $e^{-s}\neq1$
\cite{Connes1975}\cite[Theorem XVII.3.12]{TakesakiIII}.  
This does not quite imply that all one-parameter flows 
of automorphisms $\theta_s$ with $\Mod(\theta_s) = e^{-s}$ 
are conjugate, but this fact follows from the proof
of the uniqueness of the hyperfinite $\tthr_1$ factor
\cite{Haagerup1987, Connes1985}.  

The subgroup structures for the hyperfinite
factors $\rinj_0$, $\rinj_{0,1}$, and $\rinj_\infty$ 
can therefore be summarized by the following diagrams,
with inclusions going from bottom to top: for the 
hyperfinite $\ttwo_1$ factor $\rinj_0$, we have
\begin{equation*}
\tikz {
    \draw (0,0) 
    node[above]{$\aut(\rinj_0) 
    = \overline{\Int(\rinj_0)} = \uni(\rinj_0)$}
    -- (0,-1)
    node[below]{$\Int(\rinj_0) = \cnt(\rinj_0)$};
}
\end{equation*}
Next, the hyperfinite $\ttwo_\infty$ factor $\rinj_{0,1}$ subgroup
structure is
\begin{equation*}
    \tikz {
        \draw (0,0)
        node[above]{$\aut(\rinj_{0,1})$}
        -- (0,-1)
        node[fill=white]{$\overline{\Int(\rinj_{0,1})} = \uni(\rinj_{0,1})$ }
        --
        (0,-2)
        node[below]{$\Int(\rinj_{0,1}) = \cnt(\rinj_{0,1})$};
    }
\end{equation*}
Finally, the subgroup structure for the hyperfinite 
$\tthr_1$ factor $\rinj_\infty$ is 
\begin{equation*}
    \tikz{
        \draw (0,0)
        node[above]{$\aut(\rinj_\infty) 
        =\overline{\Int(\rinj_\infty)} = \uni(\rinj_\infty)$}
        -- (0,-1)
        node[fill=white]{$\cnt(\rinj_\infty)$}
        -- (0,-2)
        node[below]{$\Int(\rinj_\infty)$};
    }
\end{equation*}

\bibliography{vN-notes.bib}
\bibliographystyle{JHEP-thesis}

\end{document}